\documentclass[twocolumn,twoside,slac_two]{revtex4}
\usepackage{graphicx}
\usepackage{fancyhdr}
\newcommand{\fig}{./fig}
\newcommand{\jcite}[1]{\cite{#1}}

\newcommand{\prt}[1]{\ensuremath{\mathrm{#1}}}

\newcommand{\DZero}{{D\O}}
\newcommand{\un}[2]{\ensuremath{\mathrm{#1 \, #2}}}

\begin{document}

\title{Charm (and Beauty) Production at the Tevatron}

%

\author{Jonas Rademacker on behalf of the CDF and the \DZero\ Collaboration}
\affiliation{University of Bristol, H~H~Wills Physics Laboratory, Bristol, BS8 1TL, UK}
\begin{abstract}
 We present recent results on heavy flavour production at Tevatron
 Run~II for typically \un{\sim 1}{fb^{-1}} of analysed $p\bar{p}$ data
 at $\sqrt{s}=\un{1.96}{TeV}$. This includes results on single and
 correlated open charm and bottom cross sections, charm pair
 production kinematics, \prt{J/\psi}, \prt{\psi(2S)} and
 \prt{\chi_{cJ}} cross sections and polarisation measurements in
 \prt{J/\psi}, \prt{\psi(2S)}, \prt{\Upsilon(1S)}, and
 \prt{\Upsilon(2S)}.
\end{abstract}

\maketitle

\thispagestyle{fancy}


\section{Introduction}
 The large $b\bar{b}$ and $c\bar{c}$ cross section at
 \un{\sqrt{s}=1.96}{TeV} $p\bar{p}$ makes the Tevatron a unique place
 for the study of the production and decay of heavy flavour. Already
 in Run~I, heavy flavour production measurements challenged theory,
 finding heavy flavour production cross sections significantly higher
 than predicted by Next-to-Leading-Order (NLO) QCD calculations (see
 for example \cite{Acosta:2001rz} and \cite{Abbott:2000iv}).  The
 discrepancy between data and theory was particularly dramatic in the
 quarkonium production, where the ``Colour Singlet Model''
 leading-order QCD calculation underestimates the measured cross
 section by more than an order of magnitude (see e.g.\cite{Abe:1997jz}).

 Since the first hadroproduction measurements in Run~I, a number of
 theoretical advances have been made. Fixed-Order Next to Leading
 Logarithm (FONLL) calculations \jcite{Cacciari:1998it}, describe the
 open charm and $b$ production cross sections well. Competing models
 have been put forward that describe the observed quarkonium
 production rates and \prt{p_T} spectra well, but disagree on their
 results for quarkonium polarisation.

 There has also been dramatic experimental progress. In Run~II,
 which started in 2001, the Tevatron collides $p$ and $\bar{p}$ at
 unprecedented luminosity and energy, and the \DZero\ and the CDF
 experiment have undergone significant upgrades, many of them
 optimising the detector for flavour physics. In this paper, we
 summarise the heavy flavour production measurements in Run~II at
 \DZero\ and CDF, and compare them with theoretical predictions.

 As we will see below, heavy flavour production at hadron colliders is
 a vibrant field which is lead by experiment rather than theory. A
 particular challenge for theory is the quarkonium polarisation, for
 which we present new results from the Tevatron in this paper.

 All numbers, unless accompanied by a reference to a journal
 publication, are preliminary.

\section{The Tevatron, CDF and \DZero}
\subsection{The Tevatron Run~II}
 The Tevatron in Run~II collides protons and antiprotons at a centre
 of mass energy of \un{1.96}{TeV} with a bunch crossing every
 \un{396}{ns} at each interaction point. Some of the bunches are by
 design empty, so while the detectors have to be able to cope with
 peak rates of \un{2.5}{MHz}, the average collision rate is \un{\sim
 1.7}{MHz}.  Since the start of data taking, the Tevatron has
 delivered more than \un{3}{fb^{-1}} of integrated luminosity at each
 interaction point, and is now reaching peak luminosities of typically
 \un{2 \cdot 10^{32}}{cm^{-2}s^{-1}}, with the best runs exceeding
 \un{2.8 \cdot 10^{32}}{cm^{-2}s^{-1}}. Two general purpose detectors
 take data at the Tevatron, CDF and \DZero. Both collaborations have
 analysed approximately \un{1}{fb^{-1}} of nearly \un{3}{fb^{-1}} each
 has on tape.

\subsection{The CDF and \DZero\ Detectors}
 Both the CDF and the \DZero\ collaboration have a strong heavy
 flavour physics programme, and in the last upgrade many features have
 been added to the detectors to facilitate this programme. These
 features include precise vertexing, extended $\mu$ coverage, and
 sophisticated read-out electronics and triggers. A description of the
 \DZero\ detector can be found at \cite{Abazov:2005pn} and of the CDF
 detector at \cite{Acosta:2004yw}.

 The same detector characteristics that make \DZero\ and CDF powerful B
 physics experiments, also provide the basis for a wide-ranging charm
 physics programme. The feature that makes \DZero\ stand out as a
 heavy flavour experiment particularly is its large $\mu$ coverage
 \jcite{Abazov:2005uk}, up to $|\eta| < 2$. CDF's most distinctive
 detector element for B and charm physics is its fully hadronic
 track trigger \jcite{Ashmanskas:2003gf}.

\subsection{Triggers}
 Once every \un{396}{ns} there is a bunch crossing at each Tevatron
 interaction point, typically resulting in a very busy event; only a
 few tens of a per mill (ca $6 \cdot 10^{-4}$ at CDF) of those events
 can be written to tape and analysed in detail, the vast majority will
 be discarded instantly. The trigger is crucial, and has to operate in
 a challenging environment. There are two basic strategies to trigger
 on heavy flavour events.
\begin{itemize}
  \item Trigger on leptons ($\mu, e$). This provides a clean signature
   in an hadronic environment, where most tracks are pions. Both
   experiments use this strategy.  \DZero\ benefits particularly from
   its large $\mu$ coverage, giving large number of semileptonic and
   leptonic B and charm events, as well as charmonium and bottomium
   decays, e.g. 180k \prt{J/\psi} in the first \un{0.1}{fb^{-1}}, and
   28k \prt{B\to J/\psi K} in \un{1.6 fb^{-1}} (after selection cuts).
  \item Trigger on the long lifetimes of \prt{B} and \prt{D}
   mesons. This is very challenging in a hadronic environment because
   it requires very fast track and impact parameter reconstruction in
   very busy events, in time for the trigger decision. Both
   experiments use this strategy. But only CDF's displaced track
   trigger has enough bandwidth to trigger on fully hadronic decays
   alone, while \DZero\ have to require an additional lepton in the
   event to keep the trigger rates manageable. This high-bandwidth
   displaced track trigger gives CDF unique access to fully hadronic
   heavy flavour decays, such as 13M \prt{D^0 \to K^-\pi^+}
   events (here, and in similar expressions in the rest of the
   note, the charge-conjugate decay is always implied) , 0.3M
   \prt{D_s^+ \to \phi(K^+K^-)\pi^+} event as as well as 53k \prt{B_s
   \to D_s^{\pm} \pi^{\mp}} in \un{\sim 1}{fb^{-1}}
   \jcite{Abulencia:2006ze} (numbers after selection cuts).
\end{itemize}

\section{Cross sections}
\subsection{D and B production cross sections}
\begin{figure*}
\includegraphics[width=0.99\textwidth]{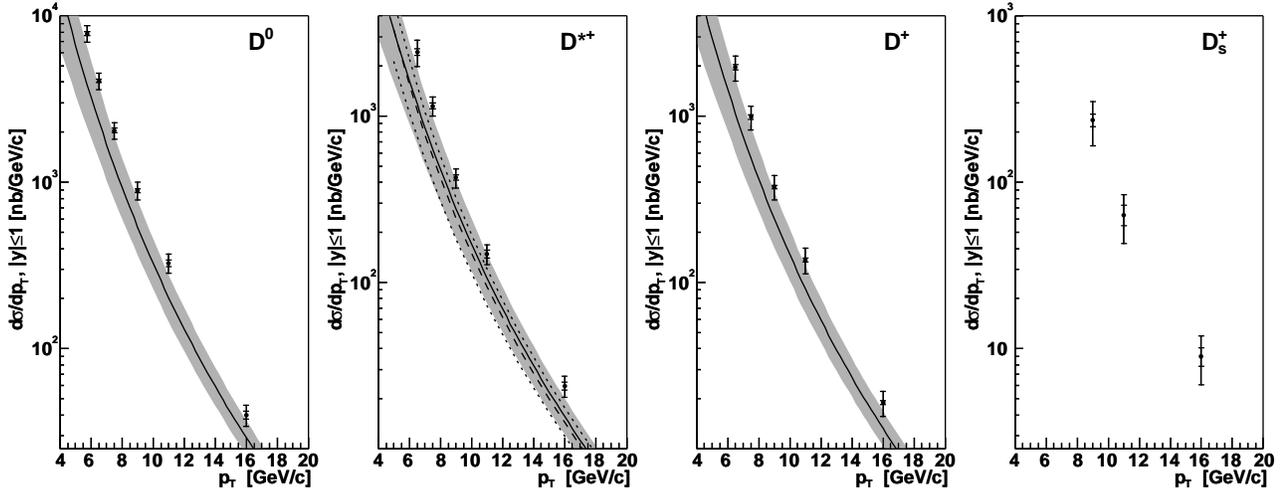}
\caption{Differential charm cross section measured in fully hadronic
  charm decays using only \un{5.8}{pb^{-1}} at CDF
  (\jcite{Acosta:2003ax}). The inner error bars represent the
  statistical uncertainty, the outer error bars represent the total
  uncertainty of the measurement, and the black line and grey band
  the central value and uncertainty of the FONLL calculation by
  \jcite{Cacciari:2003zu}. No calculation for the \prt{D_s} cross
  section was available at the time.
  \label{fig:D_xsections}}
\end{figure*}
 Amongst the earliest Tevatron Run~II results were heavy flavour cross
 section measurements. Figure \ref{fig:D_xsections} shows CDF's
 measurement of the prompt differential charm production cross section
 versus $p_T$ for \( |\eta | < 1\), for \prt{D^0}, \prt{D^+},
 \prt{D^{*+}} and \prt{D_s} mesons. Using only \un{5.8}{pb^{-1}}, less than
 0.5\% of the presently analysed dataset, the statistical error, given
 by the inner error bars, is already smaller than the systematic
 uncertainty; the combined statistical and systematic uncertainty is
 given by the outer error bars. The uncertainty on the FONLL
 calculation \jcite{Cacciari:2003zu} is given as the grey band, with
 the central value shown as a solid line. While the measured cross
 sections appear to be consistently higher than the FONLL prediction,
 the results are compatible within the (correlated) errors.

 The analyses use the reconstructed impact
 parameter of the \prt{D} to distinguish prompt \prt{D} mesons (with
 zero impact parameter) from those originating from \prt{B} decays
 (with large impact parameters).
\begin{figure}
\includegraphics[width=\columnwidth]{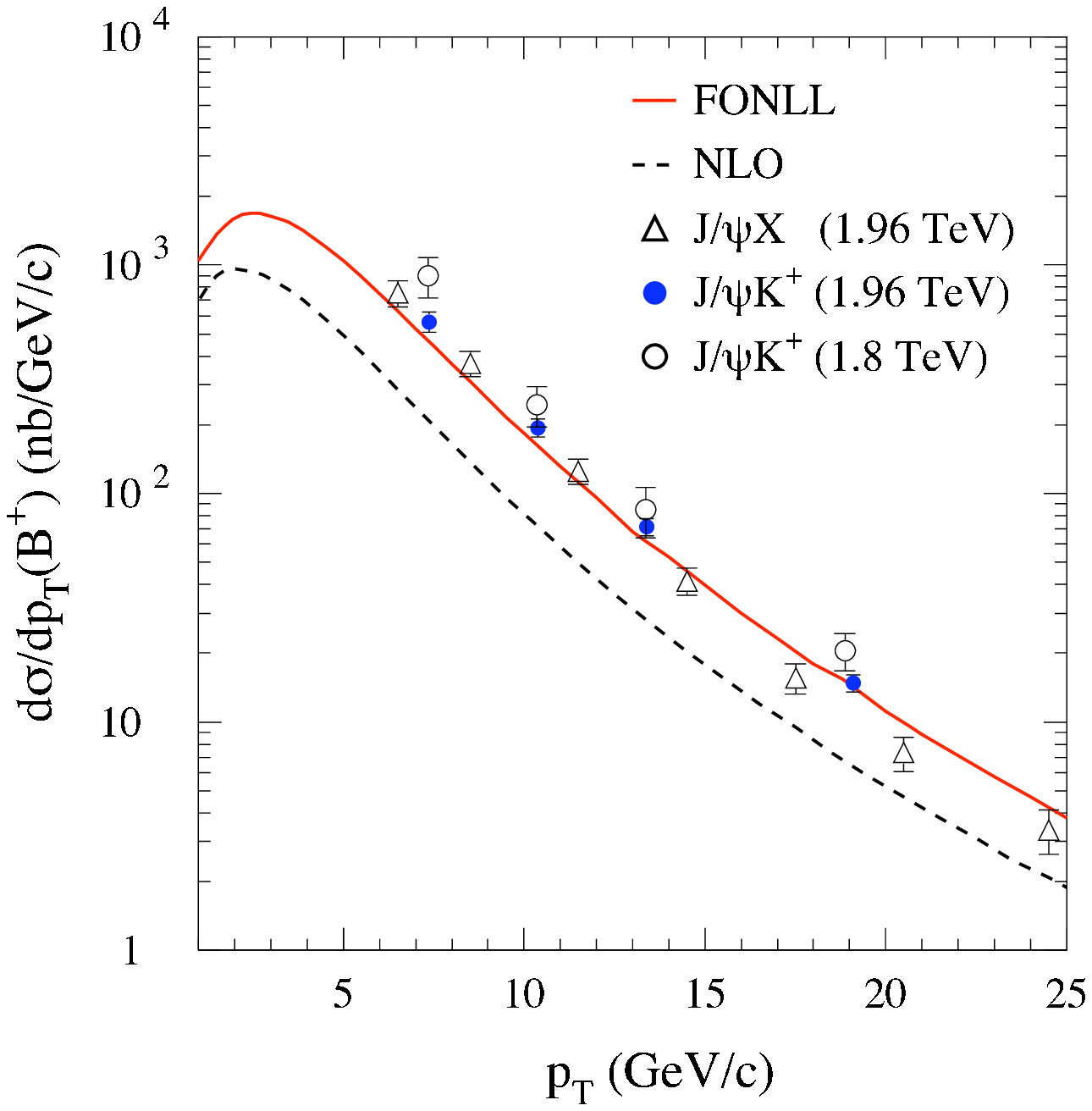}
\caption{Comparing inclusive \prt{B \to J/\psi X} \jcite{Acosta:2004yw} and exclusive
  \prt{B^+ \to J/\psi K^+} \jcite{Abulencia:2006ps}cross section
  measurements, with Run~I results \jcite{Abe:1995dv}, NLO
  \jcite{Nason:1989zy, Beenakker:1990maa}
  and FONLL \jcite{Cacciari:1998it, Cacciari:2002pa} predictions.
  \label{fig:BJpsiX}}
\end{figure}
\begin{figure}
\includegraphics[width=\columnwidth]{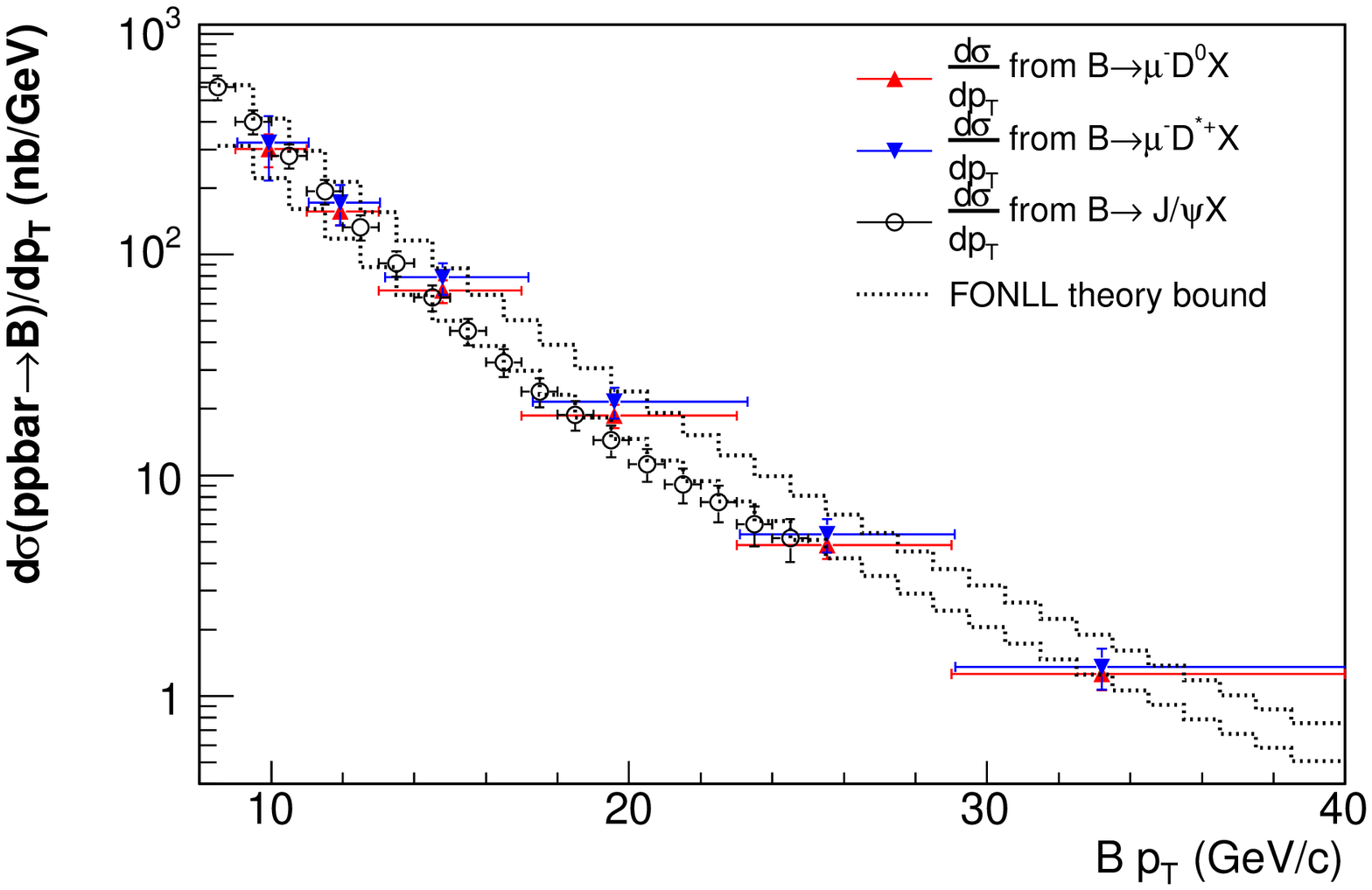}
\caption{Comparing inclusive \prt{B \to D \mu \nu X} with inclusive
  \prt{B \to J/\psi X} \jcite{Acosta:2004yw} and the FONLL calculation
  by \cite{Cacciari:2003uh}.\label{fig:BDX}}
\end{figure}

 The same technique of using either impact parameters or decay length
 has been used to measure inclusive B cross sections in \prt{B \to D^0
 \mu^+ \nu X}, \prt{B \to D^{* -} \mu^+ \nu X} and \prt{B \to J/\psi
 X} decays, where long decay lengths or large impact parameters
 identify \prt{D^0 \mu}, \prt{D^{*-} \mu} and \prt{J/\psi} originating
 from B mesons. The inclusive differential cross sections measured in
 the \prt{D^0 \mu}, \prt{D^{*-}\mu} and the \prt{J/\psi} channel are
 shown in Figures~\ref{fig:BJpsiX} and \ref{fig:BDX}. The results are
 in good agreement with FONLL calculations. The preliminary result for
 the integrated inclusive B cross section for $b$-hadrons with \un{p_t
 > 9}{GeV} and $|y| < 0.6$ is:
\begin{eqnarray*}
   \lefteqn{\sigma(\prt{p\bar{p} \to H_b}) = \un{1.34}{\mu b}}
\\ & &
       \pm \un{0.08}{\mu b (stat)}
       \un{\mbox{}^{+0.13}_{-0.14}}{\mu b(sys)}
       \pm \un{0.07}{\mu b(BR)}
\end{eqnarray*}
 for the analysis using \prt{H_b \to D^0(K^-\pi^+) \mu X}, and
\begin{eqnarray*}
   \lefteqn{\sigma(\prt{p\bar{p} \to H_b}) 
  = \un{1.47}{\mu b}}
\\ &&      \pm \un{0.18}{\mu b (stat)}
       \un{\mbox{}^{+0.17}_{-0.19}}{\mu b(sys)}
       \pm \un{0.11}{\mu b(BR)}
\end{eqnarray*}
 for the analysis using \prt{H_b \to D^{*+}(D^0(K^-\pi^+)\;\pi^+)\;
 \mu X}, where \prt{H_b} stands for a generic $b$ hadron. The last
 uncertainty is due to the uncertainty in the branching fractions of
 the specific final states of the \prt{D^0} and \prt{D^*} being
 investigated. The result is in good agreement with the FONLL value of
 \un{1.39^{+0.49}_{-0.34}}{\mu b} \jcite{Cacciari:2003uh}.

 CDF performed a measurement of the exclusive \prt{B^+ \to J/\psi
 K^+} measurement, using \un{0.74}{fb^{-1}}, finding \( \sigma\left(p_t
 > \un{6}{GeV}, |y| < 1\right) = \left( 2.65 \pm 0.12(stat) \pm
 0.21(sys) \right) \mathrm{\mu b} \) \jcite{Abulencia:2006ps}.  The
 differential cross sections for the exclusive measurement can be seen
 in Fig \ref{fig:BJpsiX} together with the \prt{B \to J/\psi X}
 inclusive results. All measurements disagree significantly with NLO
 calculations. The agreement with FONLL is however very good.

\subsection{Correlated $\bf b\bar{b}$ and $\bf c\bar{c}$ cross sections}

\begin{table}
\begin{tabular}{c|*{4}{c}}
\multicolumn{1}{c}{} & \multicolumn{4}{c}{Ratio:
  measurement/NLO-prediction in 2006}\\
\multicolumn{1}{c}{$p_t^{\mathrm{min}}=$} & \multicolumn{1}{c}{6-7} GeV & 10GeV & 15GeV & $\sim 20$ GeV\\\hline
channel &&&&\\
$b + \bar{b}$ jets &   &   & $1.2\pm 0.3$ & $1.0 \pm 0.3$\\
$\mu + b$ jet  &   & $1.5\pm 0.2$  & & \\
$\mu^+ + \mu^-$    & $3.0\pm 0.6$ & & &\\
$\mu^+ + \mu^-$    & $2.3\pm 0.8$ & & &
\end{tabular}
\caption{Ratios of measurements and predictions for correlated
  $b\bar{b}$ cross section measurements, presented by F.~Happacher at
  DIS~2006~\cite{Happacher:2006im}. New (2007) results from CDF in the
  \prt{\mu^+\mu^-} channel are presented in here.
  (Table taken from \cite{Happacher:2006im}, slightly modified.)
  \label{tab:correllated_bb_pastResults}
  }
\end{table}
 For correlated $b\bar{b}$ and $c\bar{c}$ cross sections, i.e. cross
 sections where both the quark and the antiquark are within a certain,
 central rapidity range, higher order terms are expected to be smaller
 and consequently NLO calculations are expected to describe the data
 better \jcite{Acosta:2003nn, Happacher:2006im}. In the past, Tevatron
 results on correlated $b\bar{b}$ cross sections have been
 inconclusive. Table~\ref{tab:correllated_bb_pastResults} shows the
 status presented by \cite{Happacher:2006im} at DIS~2006. Especially
 in the $\mu^+ \mu^-$ channel, results have been at odds with NLO
 predictions. This year (2007) CDF have presented a new measurement of
 the correlated $b\bar{b}$ and $c\bar{c}$ cross section in the $\mu^+
 \mu^-$ channel, based on \un{0.74}{fb^{-1}}. The study uses $\mu^+
 \mu^-$ pairs with transverse momentum $p_T > \un{3}{GeV}$,
 pseudorapidity $|\eta| < 0.7$ and invariant mass $m_{\mu \mu} \in
 \un{[5,80]}{GeV}$. This corresponds to $b\bar{b}$ pairs with $p_T \ge
 \un{2}{GeV}$ and a rapidity $|y| \le 1.3$.
\begin{figure}
\begin{tabular}{cc}
 \includegraphics[width=0.99\columnwidth]{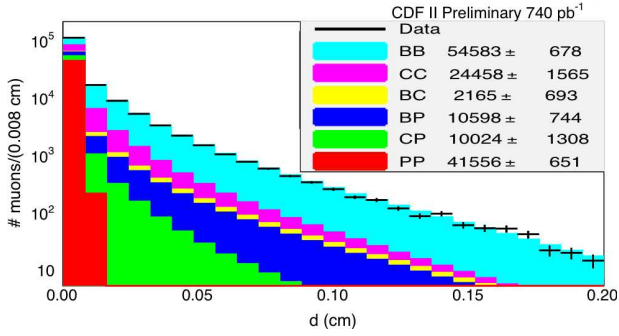}
\end{tabular}
\caption{Impact parameter ($d$) distribution of $\mu^+ \mu^-$ pairs.
  Although only the projection onto 1 dimension is shown, the fit is
  performed on the full 2-D distribution of $\mu^+$ and $\mu^-$ impact
  parameters. The different letter combinations that can be found in
  the legend distinguish muon pairs from B (``BB''), from charm
  (``CC''), and prompt pairs (``PP'') and various mixtures, such as
  ``BP'' where one muon is prompt and the other from a B
  decay. \label{fig:mumuIP}}
\end{figure}
 Prompt $\mu$ are separated from $\mu$ created in charm decays, and
 those from $\mu$ created in B decays, using the impact parameters of
 each muon in the $\mu^+ \mu^-$ pair. The 1-D projection of the impact
 parameter distribution, and fit, is shown in Fig~\ref{fig:mumuIP}.

 The preliminary result for the correlated $b\bar{b}$ production cross
 section, where each $b$ quark decays to a $\mu$, is
\[
 \sigma_{b \to \mu, \bar{b} \to \mu}
 = \un{\left( 1549 \pm 133 \right)}{pb}.
\]
 and for $c\bar{c}$ production where each charm
 quark decays to a $\mu$ is
\[
 \sigma_{c \to \mu, \bar{c} \to \mu} = \un{\left( 624 \pm 104\right)}{pb}
\]
 Translating the $b\to \mu,\;\; \bar{b}\to \mu$ cross section to an
 inclusive $b\bar{b}$ cross section, independent of the final state:
\begin{eqnarray*}
\lefteqn{\sigma_{b\bar{b}}\left(p_T \ge \un{6}{GeV}, |y| \le 1\right)
 =}\\
 && \un{\left( 1618 \pm 148 \pm [\sim 400\; \mathrm{fragmentation}]
 \right)}{nb}
\end{eqnarray*}
 where the dominant error comes from the uncertainty in the
 fragmentation function, i.e. the fraction of $b$ quarks that decay to
 $\mu$. Different values for the Peterson fragmentation parameter
 $\epsilon$ \jcite{Peterson:1982ak} result in significantly different
 results, especially for the correlated $c\bar{c}$ cross
 section.
 \begin{table}
 \begin{center}
  \begin{tabular}{|*{3}{c|}}
\hline
   & $\epsilon = 0.006$ & $\epsilon = 0.002$ \\\hline
\( \displaystyle \large
\frac{\sigma_{b\to \mu\; \bar{b}\to \mu}^{\mathrm{measured}}
  }{\sigma_{b \to \mu\; \bar{b} \to \mu}^{\mathrm{NLO}}}
\)
&
  $ 1.2 \pm 0.2$
&
  $ 1.0 \pm 0.2$
\\
 \(\displaystyle \large
\frac{\sigma_{c\to \mu\; \bar{c}\to \mu}^{\mathrm{measured}}
  }{\sigma_{c \to \mu\; \bar{c} \to \mu}^{\mathrm{NLO}}}
\)
&
 $ 2.7 \pm 0.6$
&
 $ 1.6 \pm 0.4$
\\\hline
  \end{tabular}
\caption{Measured correlated $b\bar{b}$ and $c\bar{c}$ cross sections in
  \un{0.74}{fb^{-1}} at CDF, divided by NLO predictions, for different
  assumptions for the Peterson fragmentation parameter
  $\epsilon$ \jcite{Peterson:1982ak}.\label{tab:corrXRatios}}
 \end{center}
 \end{table}
 This can be seen in Table \ref{tab:corrXRatios}, where the ratio of
 the measured cross sections and the NLO predictions is given for two
 values of the Peterson fragmentation parameter $\epsilon$. The value
 traditionally used, $\epsilon = 0.006$, is obtained from fits to
 $e^+, e^-$ data \jcite{Chrin:1987yd}. However, \cite{Cacciari:2002pa}
 point out that these calculations, made on the basis of LO parton
 level cross sections evaluated with the leading-log approximation of
 the parton shower event generator, cannot be consistently used with
 exact NLO calculations, and find that a more suitable value of the
 Peterson fragmentation parameter would be $\epsilon = 0.002$. For
 $b\bar{b}$ the results for both values of $\epsilon$ are now in good
 agreement with NLO predictions, in contrast to previous
 measurements. For $c\bar{c}$ the agreement between measurement and
 NLO prediction depends crucially on the choice of $\epsilon$, and no
 firm statements regarding the compatibility between CDF's correlated
 $c\bar{c}$ cross section measurement and NLO calculations can be made
 until the issue of a suitable fragmentation parameter is resolved.

\section{Charm Pair Cross Section}
\label{sec:pairX}

\begin{figure}
\begin{tabular}{|cc|}
\hline
 \parbox{0.52\columnwidth}{
\includegraphics[width=0.51\columnwidth]{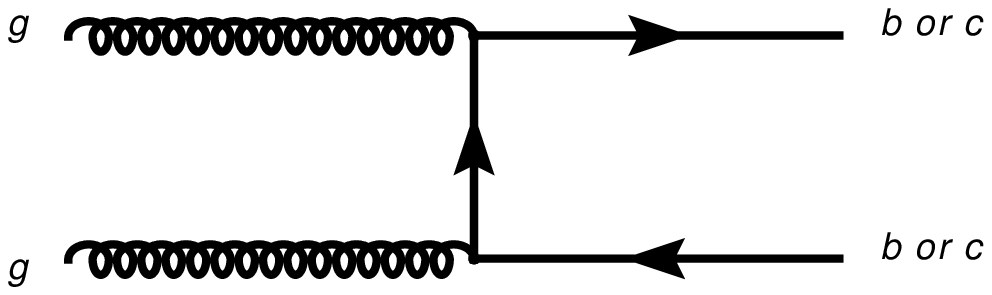}
\\
\includegraphics[width=0.51\columnwidth]{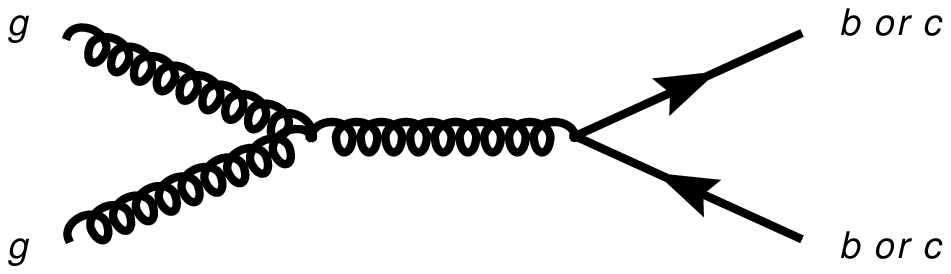}
}
&\parbox{0.44\columnwidth}{
 ``Flavour Creation'' - favours large $\Delta\phi$,
   back-to-back production.
}
\\\hline
 \parbox{0.52\columnwidth}{
\includegraphics[width=0.51\columnwidth]{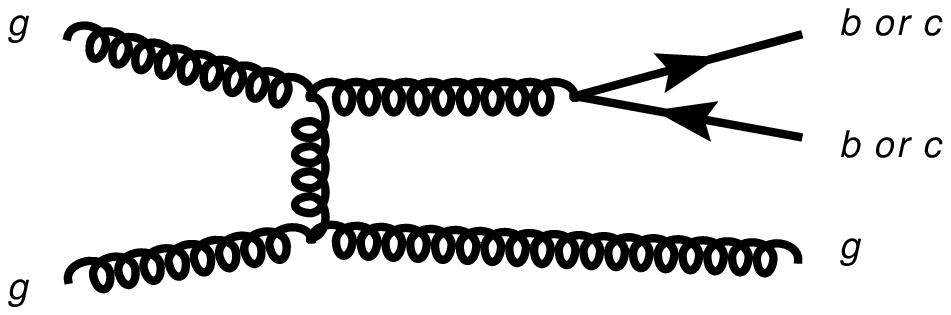}
}
&\parbox{0.44\columnwidth}{
 ``Gluon Splitting'' - favours small $\Delta\phi$,
     collinear production.
}
\\\hline
 \parbox{0.52\columnwidth}{
\includegraphics[width=0.51\columnwidth]{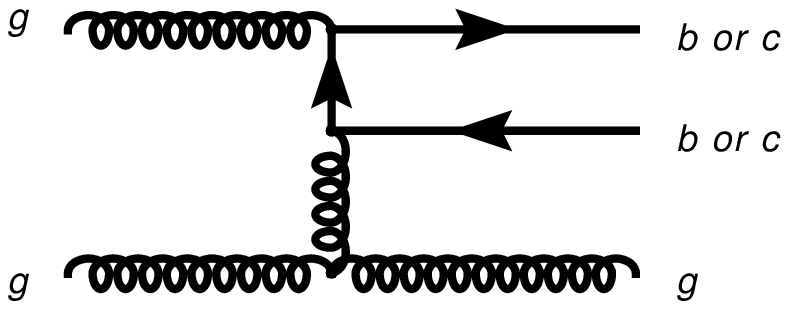}
}
&\parbox{0.44\columnwidth}{
 ``Flavour Excitation'' - characterised by relatively large $\Delta\eta$.
}
\\\hline
\end{tabular}
\caption{Heavy flavour production mechanisms in \prt{p\bar{p}}
  collisions. Different mechanisms lead to different
  kinematics.\label{fig:productionMechanisms}}
\end{figure}
 Different production mechanisms for heavy flavour quark-antiquark
 pairs lead to different kinematic distributions. The leading
 production mechanism, and their kinematic signature, are depicted in
 Fig~\ref{fig:productionMechanisms}. The lowest order diagrams
 (``Flavour Creation'') favours back-to-back production of the
 quark-antiquark pair, while ``Gluon Splitting'' favours collinear
 production. Measuring the angular distribution in charm pair
 production therefore gives clues about the $c\bar{c}$ production
 mechanism.

\begin{figure}
\parbox{0.99\columnwidth}{
 \includegraphics[width=0.98\columnwidth]{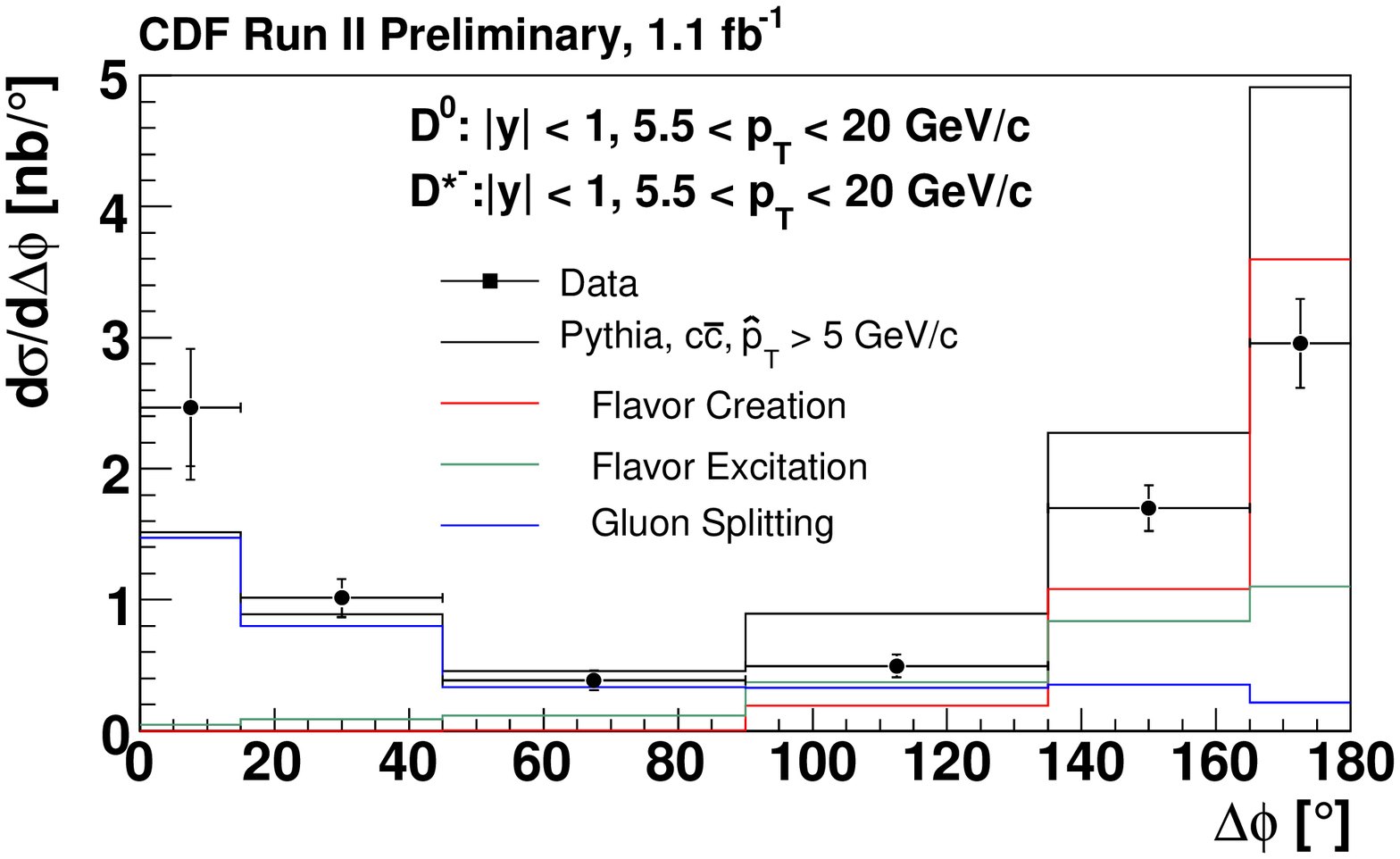}
\\
 \includegraphics[width=0.98\columnwidth]{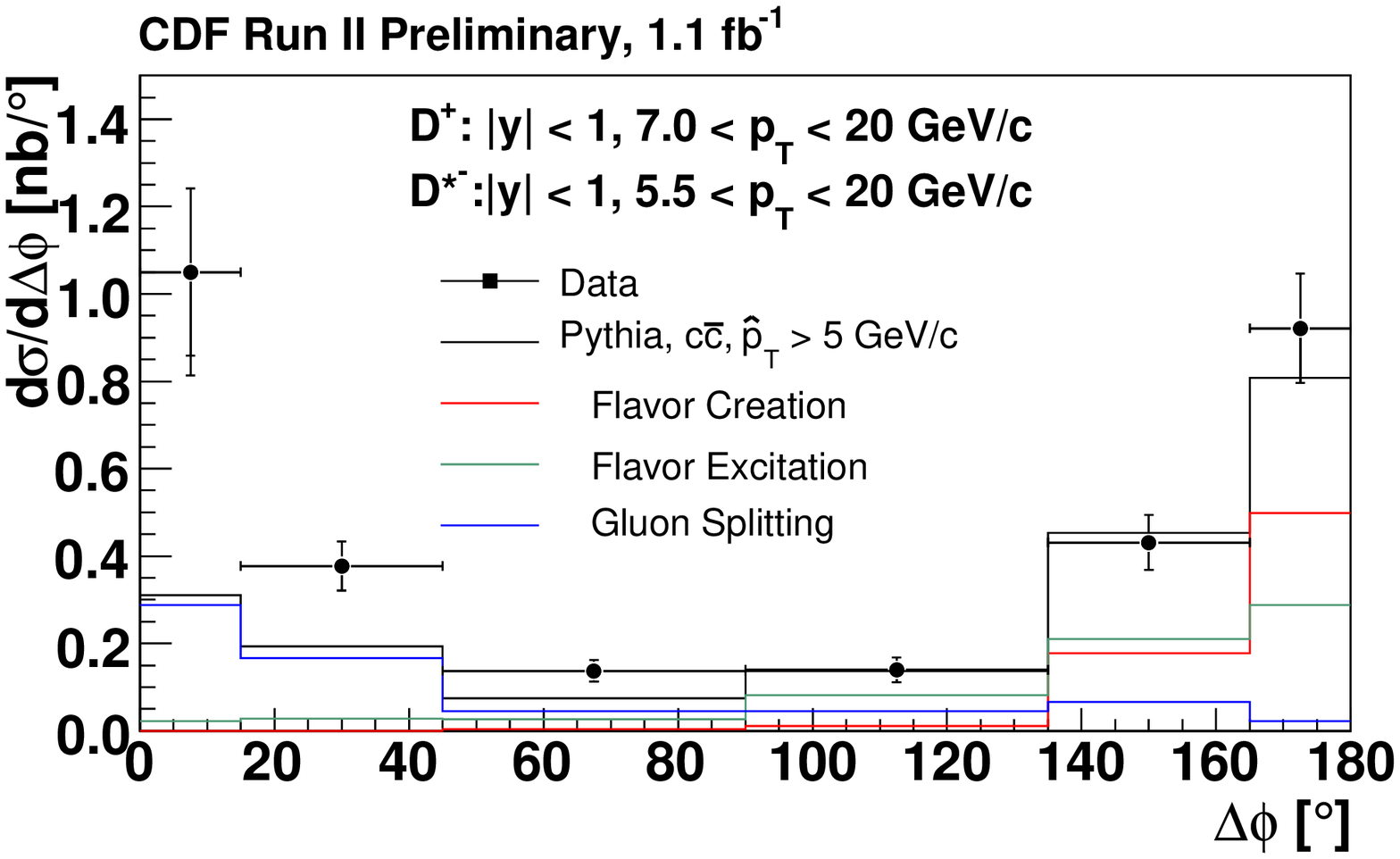}
}
\caption{\prt{D^0}, \prt{D^*} pair cross section and \prt{D^+},
  \prt{D^*} pair cross section as a function of the azimuthal angle
  between the to charm mesons (dots with error bars). The black line
  shows the result from the Pythia event generator
  \jcite{Sjostrand:2001yu}, run with leading order matrix elements +
  parton shower, tune A \jcite{Field:2002vt, Field:2005sa,
  Field:2002_CDFnote, Field:2002_web}. The coloured lines show the
  contribution to the simulated events from different production
  mechanisms.
\label{fig:DzeroPair}}
\end{figure}
 Figure \ref{fig:DzeroPair} shows the \prt{D^0}, \prt{D^{*-}} pair
 cross section and the \prt{D^+}, \prt{D^{*-}} pair cross section as a
 function of the angle between the two charm mesons, for the kinematic
 range $|y^{D^0}| < 1$, $p_T^{D^0} \in \un{[5.5, 20]}{GeV}$ and
 $|y^{D^*}| < 1$, $p_T^{D^*} \in \un{[5.5, 20]}{GeV}$. Using a
 \prt{D^*} in the reconstruction ensures clean data samples. It can be
 seen that in both cases, collinear production is approximately of the
 same size as back-to-back production.  The figure also shows the
 result from the Pythia event generator, configured to run with
 leading order matrix elements plus parton shower (Tune A). The
 total pair production cross sections agree well between the Pythia
 simulation and the data. However, the simulation over-estimates
 back-to-back production and under-estimates collinear production.

\section{Charmonium, Bottomium}

 Producing a colour-neutral $J^P=1^-$ state directly by gluon-gluon
 fusion is not possible. The simplest solution to the problem is to
 produce a colour-charged $c\bar{c}$ (or $b\bar{b}$) pair with gluon
 fusion and ``bleach'' it by radiating off a gluon.
\begin{figure}
\begin{center}
\begin{tabular}{c|c}
(a) Colour & (b) QCD with 
\\
Singlet & higher-order terms
\\
\includegraphics[height=0.3\columnwidth]{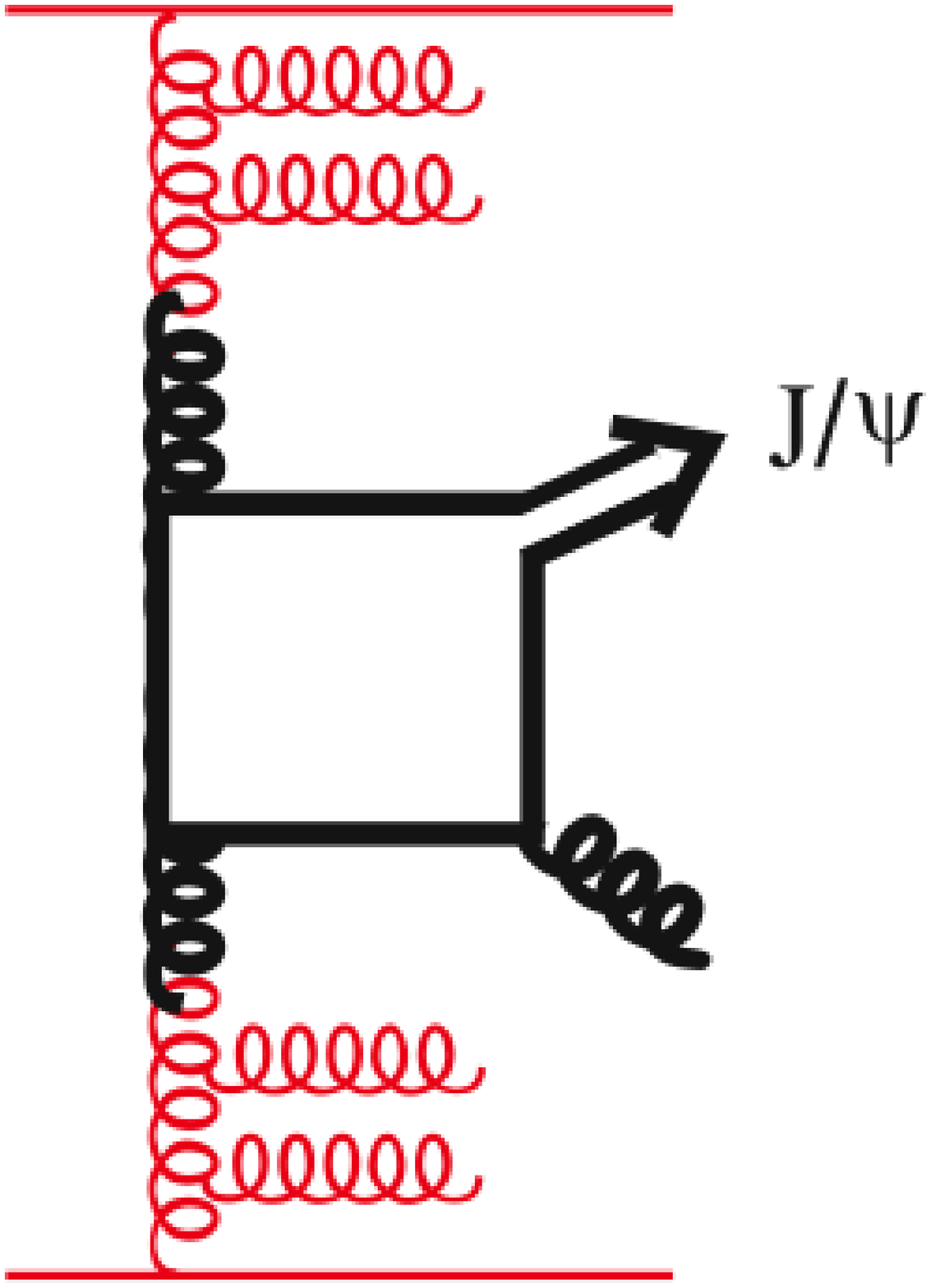}
&
\mbox{
\includegraphics[height=0.3\columnwidth]{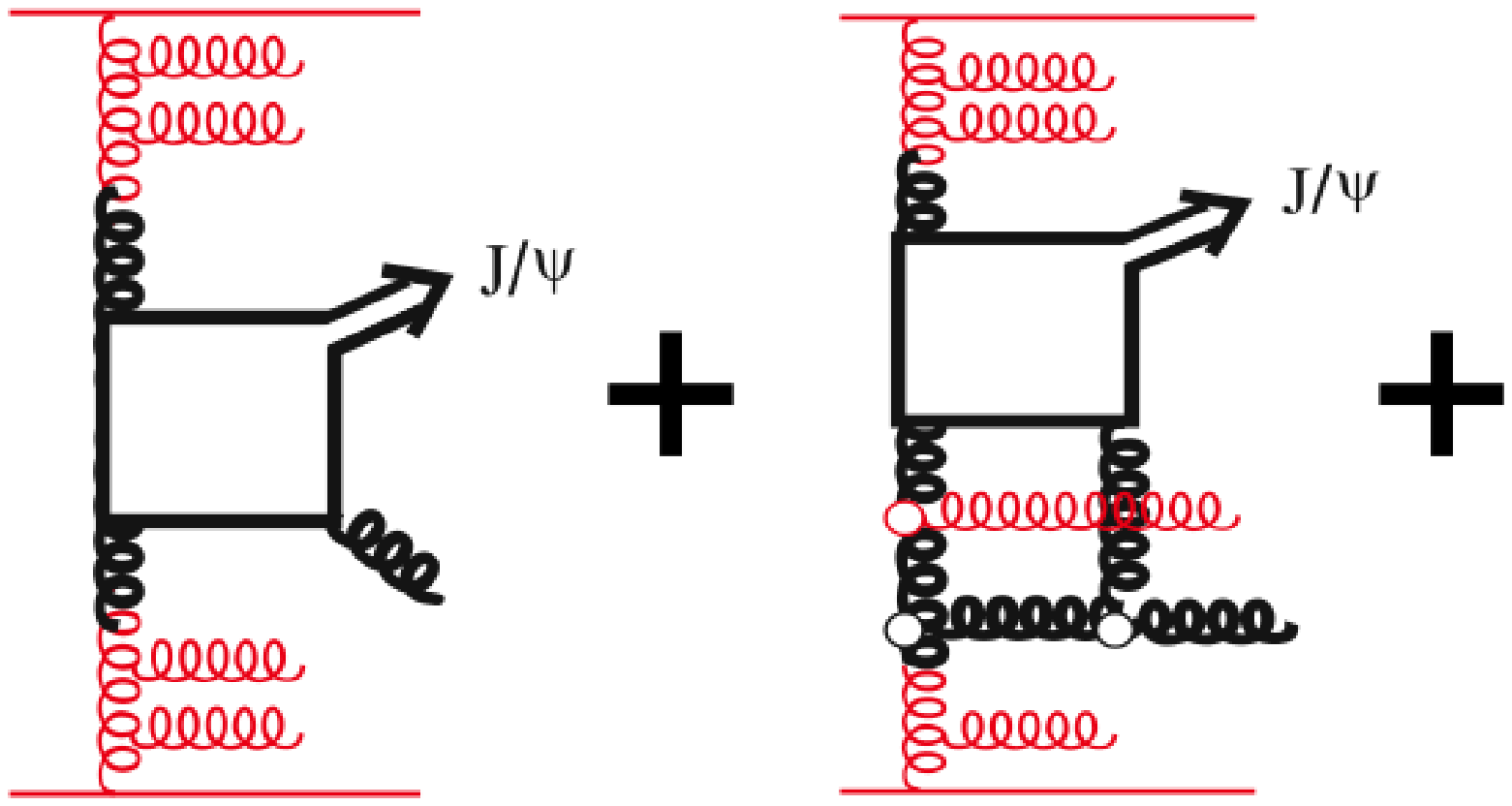}
{\Huge$\mathbf{\cdots}$}
}
\end{tabular}
\caption{``Colour singlet'' lowest order QCD (left), and QCD with
  higher order terms (right). The hard scattering process is shown in
  bold. The higher order terms considered are those where the
  ``bleaching gluon'' is absorbed by one of the parton
  spectators \jcite{Khoze:2004eu}.\label{fig:oniumProductionDiagrams}}
\end{center}
\end{figure}
  The ``colour-singlet'' mechanism,
 illustrated in Fig~\ref{fig:oniumProductionDiagrams} (a), was first
 used to describe quarkonium production \jcite{Baier:1983va,
 Guberina:1980dc}, but dramatically fails to describe the observed
 data at the Tevatron, under-estimating the \prt{J/\psi} and
 \prt{\psi^{\prime}} production in Run~I by more than an order of
 magnitude (see e.g. \cite{Braaten:1994xb}). The discussion about a
 solution to this problem is dominated by two approaches:
\begin{itemize}
  \item The ``Colour Octet'' mechanism proposed by
    \cite{Braaten:1994vv} is an effective field theory model based on
    Non-Relativistic QCD (NRQCD). It combines results from
    \cite{Bodwin:1994jh} and elements of the Colour Evaporation model
    (CEM). The CEM was originally proposed in~\cite{Fritzsch:1977ay,
    Halzen:1977rs}; more recent discussions can be found in
    \cite{Amundson:1995em, Amundson:1996qr}. In the CEM the
    \prt{J/\psi} is essentially formed in a coloured state, so no
    ``bleaching gluon'' is needed. Then the colour ``evaporates'' in
    the emission of soft gluons.
  \item \cite{Khoze:2004eu} perform a calculation based on full,
    relativistic QCD, adding higher-order terms to the colour singlet
    term of the type shown in
    Fig~\ref{fig:oniumProductionDiagrams}(b), where the ``bleaching
    gluon'' is absorbed by a spectator parton. Although each additional
    term is small, there is such a large number of such diagrams that
    the sum of them does indeed make a large enough contribution to
    account for the order-of-magnitude difference between the observed
    cross sections and those predicted by the colour-singlet model.
\end{itemize}
  Both NRQCD colour-octet and higher-order perturbative QCD describe
  the observed \prt{J/\psi} and \prt{\psi^{\prime}} cross sections and
  $p_t$ spectra well (see for example \cite{Braaten:1994vv,
  Khoze:2004eu, Hagler:2000eu}), where the NRQCD approach has a number
  of adjustable hadronisation parameters that allow a certain level of
  tuning. However, NRQCD makes a firm prediction that the \prt{J/\psi}
  should be transversely polarised \jcite{Braaten:1999qk} while
  perturbative QCD predicts a longitudinal polarisation of the
  \prt{J/\psi} \jcite{Khoze:2004eu}. 

  The $k_t$ factorisation approach \jcite{Catani:1990xk,
  Catani:1990eg, Collins:1991ty, Camici:1996st, Camici:1997ta,
  Ryskin:1999yq, Hagler:2000dda, Baranov:2002cf, Baranov:2007ay}, can
  be combined with the colour singlet and the colour octet
  mechanism. In contrast to the usual collinear approach, $k_T$
  factorisation takes the non-vanishing transverse momentum of the
  interacting gluons into account when calculating the hadronic matrix
  element. Especially when combined with the colour-octet mechanism,
  it describes the production cross sections well, although there is a
  number of adjustable parameters that are not yet fixed by
  experimental data \jcite{Hagler:2000dda, Baranov:2002cf}. In
  contrast to the usual NRQCD colour-octet mechanism, $k_T$
  factorisation predicts a longitudinal polarisation of the quarkonium
  that increases with $p_T$.

  For a more complete and detailed review of the theory of quarkonium
  production at hadron colliders, including comparison to data, see
  \cite{Lansberg:2006dh} and references therein.
\subsection{The $\mathbf{\psi(2S)}$ Cross Section}
\begin{figure}
\includegraphics[width=0.9\columnwidth]{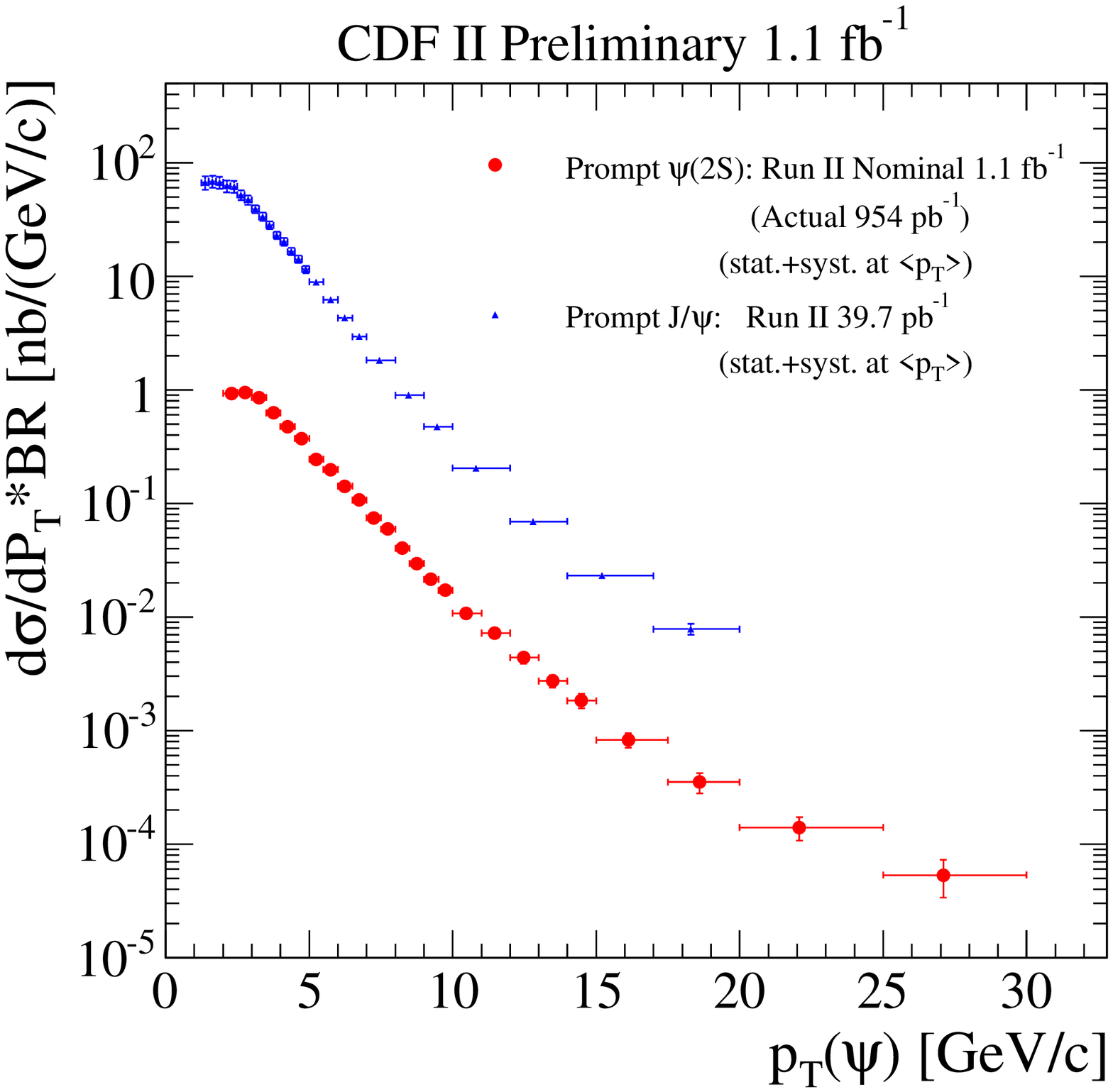}
 \caption{Differential cross section of prompt \prt{\psi(2S)} cross
 section in \un{\sim 1}{fb^{-1}}, with an earlier result for the
 \prt{J/\psi} cross section \jcite{Acosta:2004yw}
 superimposed.\label{fig:JpsiPsi2X}}
\end{figure}
 The differential cross section of prompt \prt{\psi(2S)} from \un{\sim
 1}{fb^{-1}} at CDF Run~II and, for comparison the differential cross section
 of prompt \prt{J/\psi} from \un{39.7}{pb} at CDF Run~II is shown in
 Fig~\ref{fig:JpsiPsi2X}. The preliminary result for the integrated
 cross section of \(p\bar{p} \to \psi(2S)X\) at
 $\sqrt{s}=\un{1.96}{TeV}$, with a subsequent decay \prt{\psi(2S) \to
 \mu^+ \mu^-}, in the kinematic range
 \(|y_{\psi(2S)}| < 0.6, p_T > \un{2}{GeV} \), is
\begin{eqnarray*}
\lefteqn{
 \sigma(p\bar{p} \to \prt{\psi(2S)X}, {|y_{\prt{\psi(2S)}}| < 0.6, p_T >
 \un{2}{GeV}})}
\\
 &\times&  Br(\prt{\psi(2S) \to \mu^+ \mu^-})
\\
  &&= \left(2.60 \pm 0.05(stat) \mbox{}^{+0.19}_{-0.18}(syst)\right)
 \mathrm{nb}
\end{eqnarray*}

\subsection{Measurement of charmonium polarisation}
  The $J/\psi$ can have three polarisation states, two transverse and
  one longitudinal. In \prt{J/\psi \to \mu^+\mu^-} decays, the
  transverse and longitudinal polarisation states can be disentangled
  by measuring the angle $\theta^*$ between the $J/\psi$ and the
  $\mu^+$ in the the $J/\psi$ restframe. It is useful to define the
  parameter $\alpha$ in terms of the cross section of transversely
  polarised $J/\psi$, $\sigma_T$, and the cross section of
  longitudinally polarised $J/\psi$, $\sigma_L$, as:
\begin{equation}
 \alpha \equiv \frac{\sigma_T - 2\sigma_L}{\sigma_T + 2\sigma_L}
\end{equation}
  If the one longitudinal and the two transverse polarisation states
  are all equally populated, one would measure $\alpha=0$; for longitudinal
  polarisation, $\alpha < 0$, and for transverse polarisation, $\alpha
  > 0$. The parameter $\alpha$ can be extracted from the distribution
  of events as a function of $\cos\theta^*$:
\begin{equation}
 \frac{dN}{d(\cos\theta^*)} \propto 1 + \alpha \cos^2\theta^*
\end{equation}

\begin{figure}
 \includegraphics[width=0.9\columnwidth]{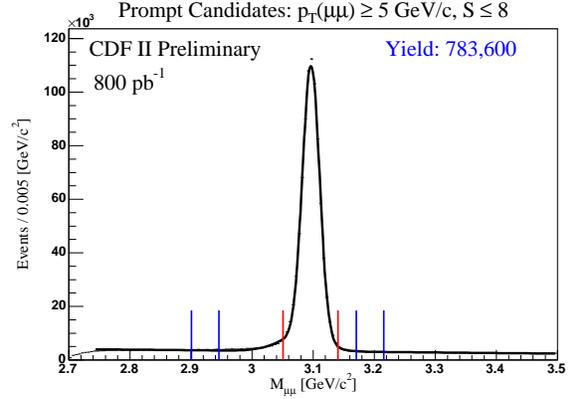}
\caption{Invariant mass of $\mu^+ \mu^-$ pairs at CDF (after selection
  cuts). CDF finds $0.78M$ prompt \prt{J/\psi} in
  \un{0.8}{fb^{-1}} \cite{Abulencia:2006_JpsiPol}.\label{fig:JpsiMass}}
\end{figure}
\begin{figure}
\begin{tabular}{ccc}
\includegraphics[width=0.3\columnwidth]{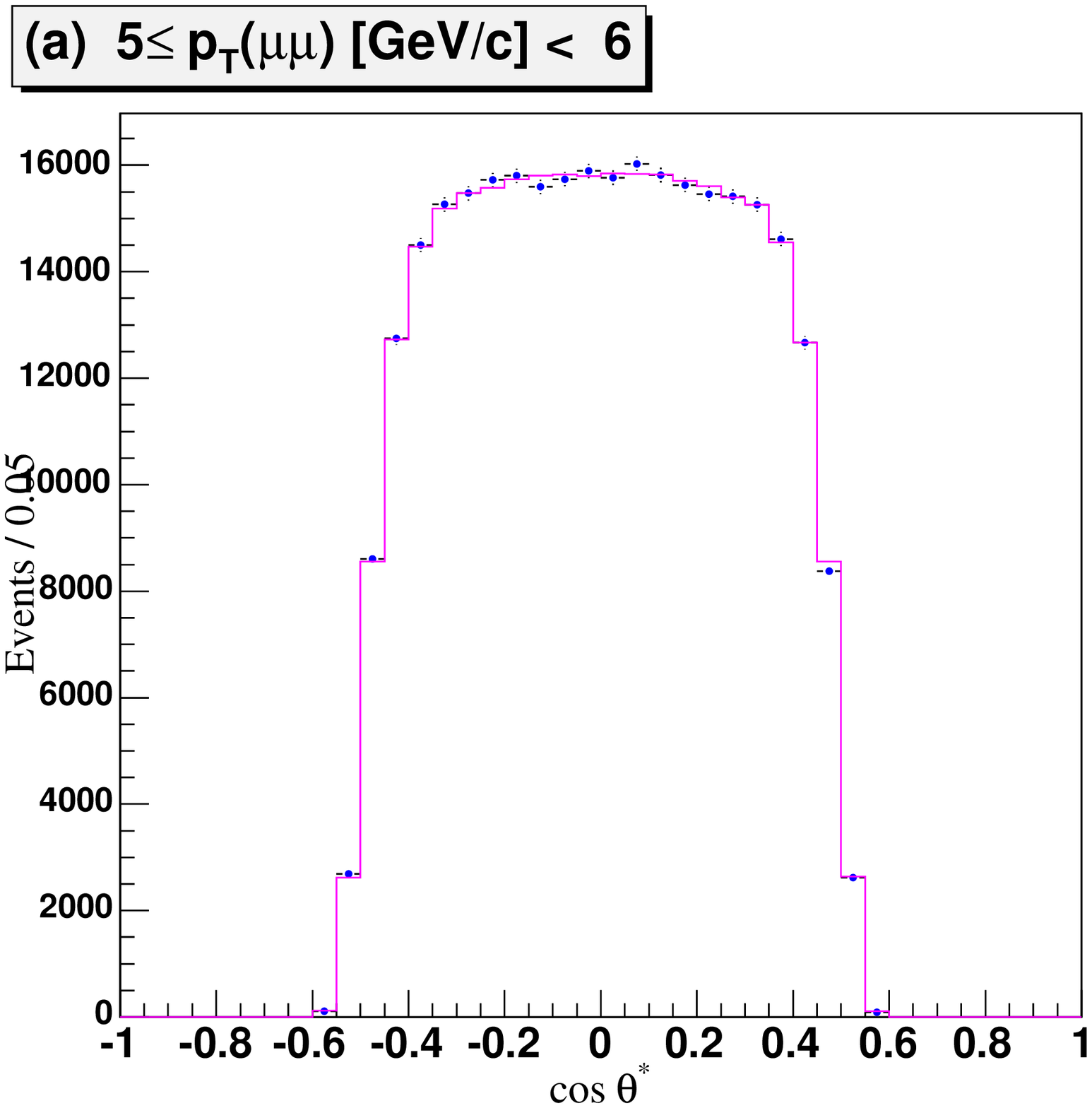}
&
\includegraphics[width=0.3\columnwidth]{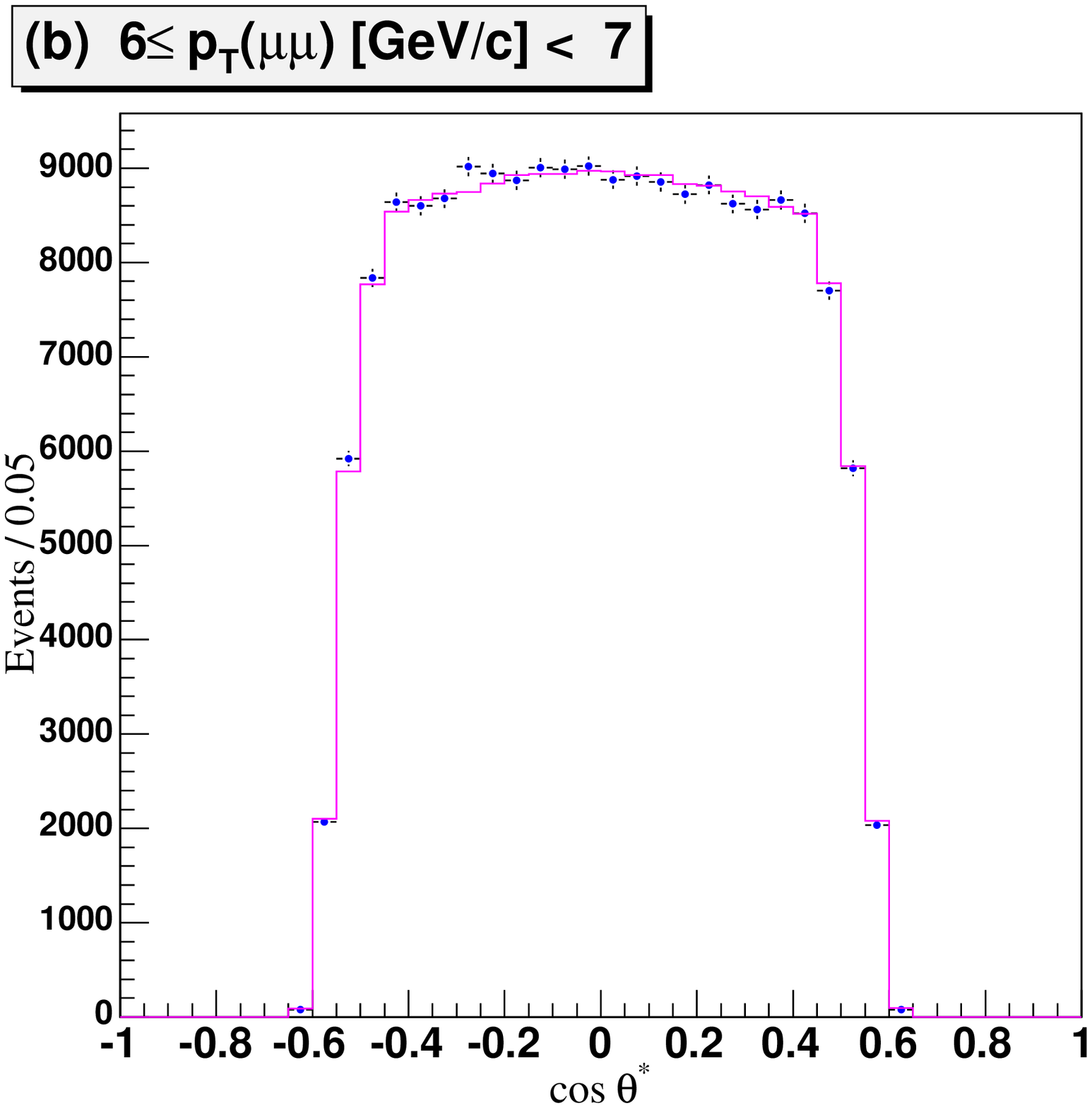}
&
\includegraphics[width=0.3\columnwidth]{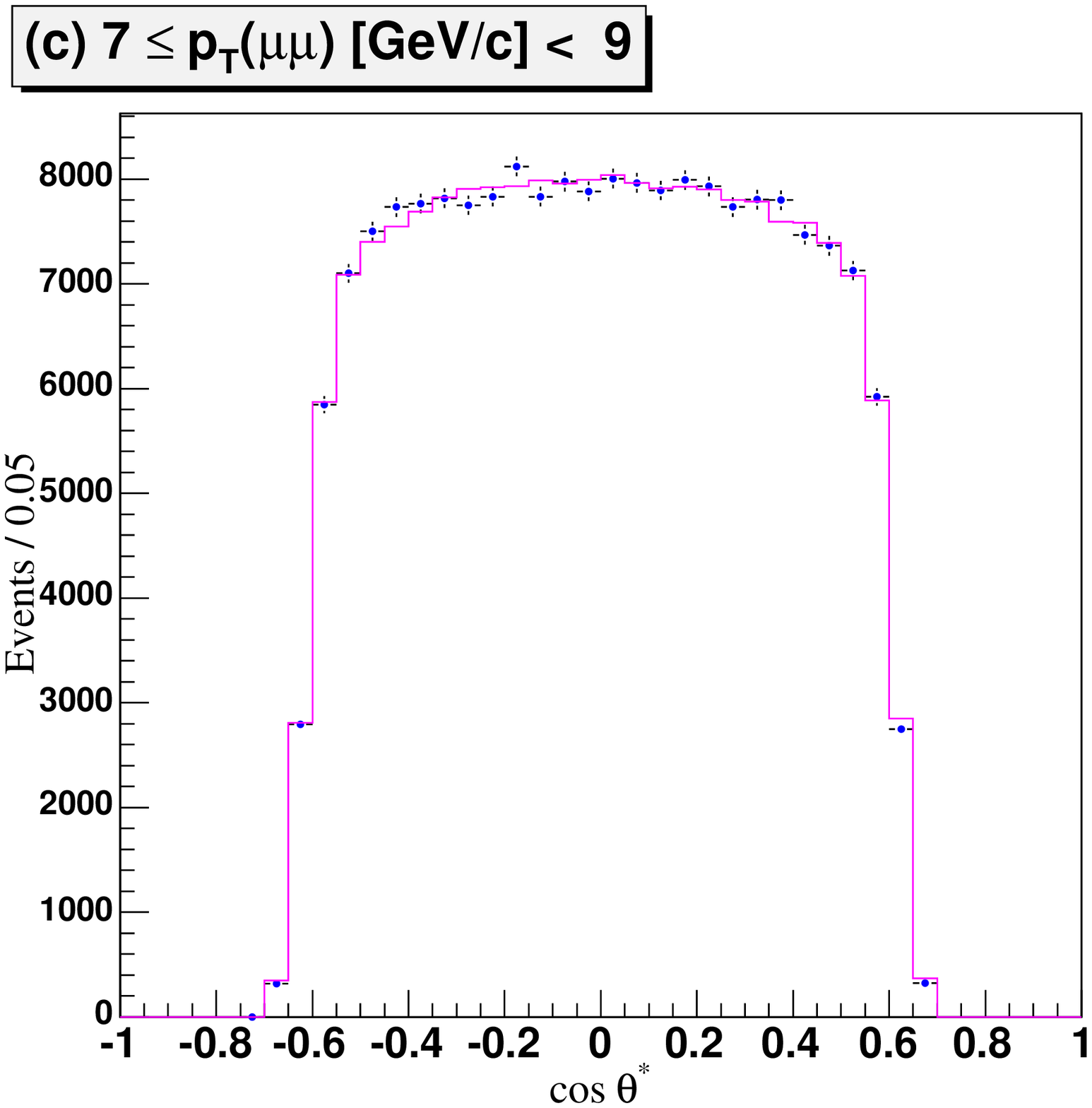}
\\
\includegraphics[width=0.3\columnwidth]{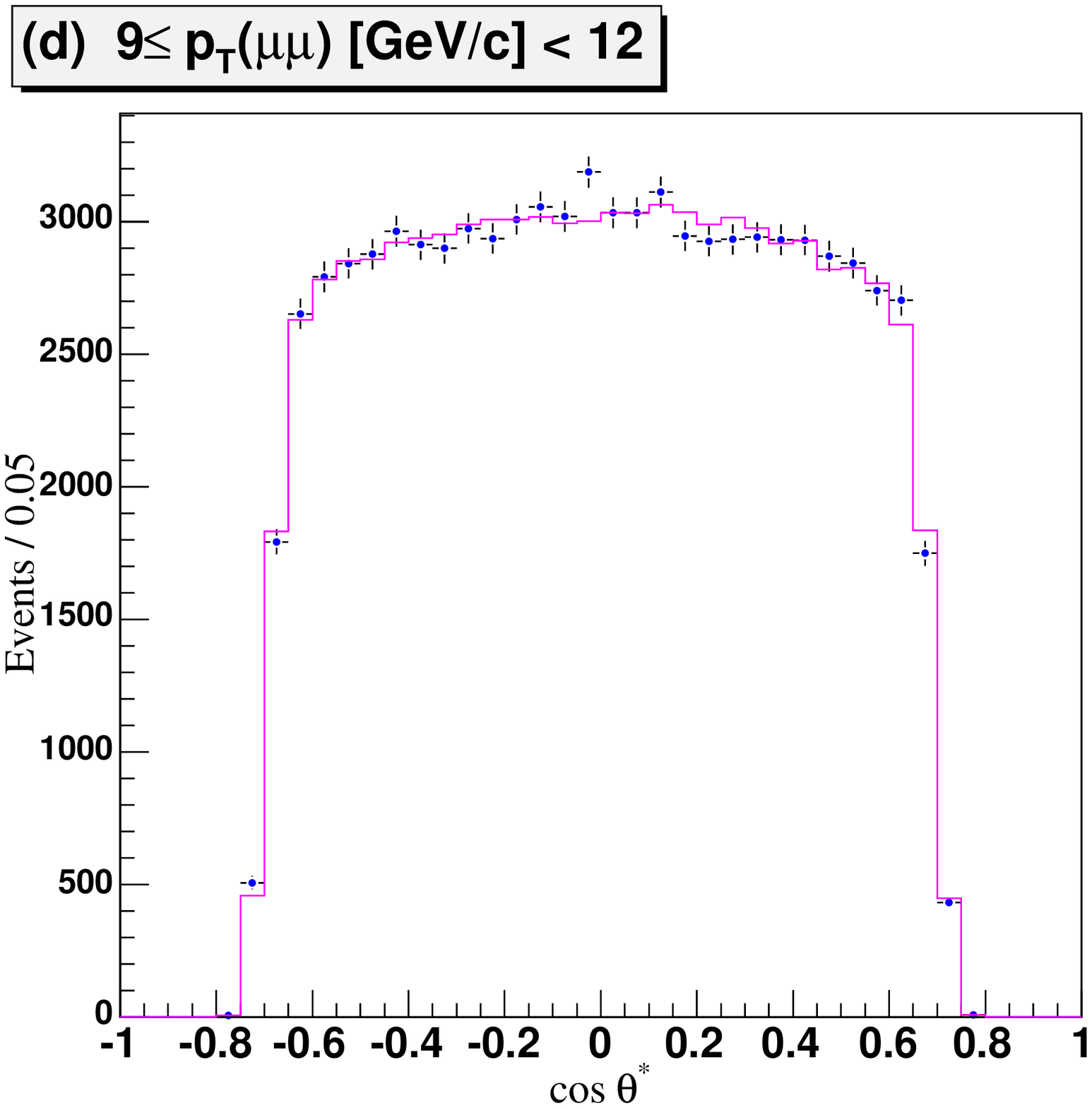}
&
\includegraphics[width=0.3\columnwidth]{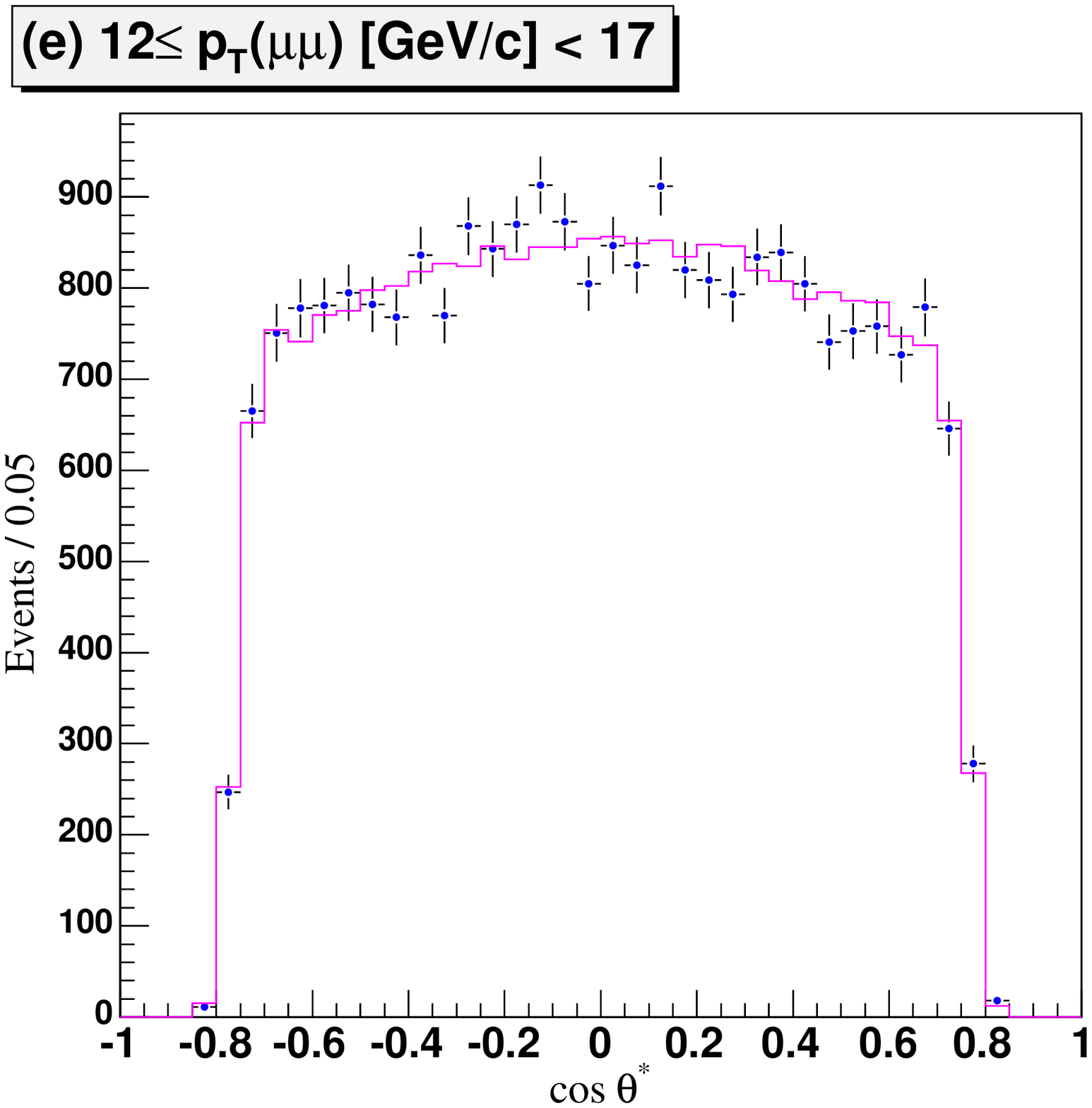}
&
\includegraphics[width=0.3\columnwidth]{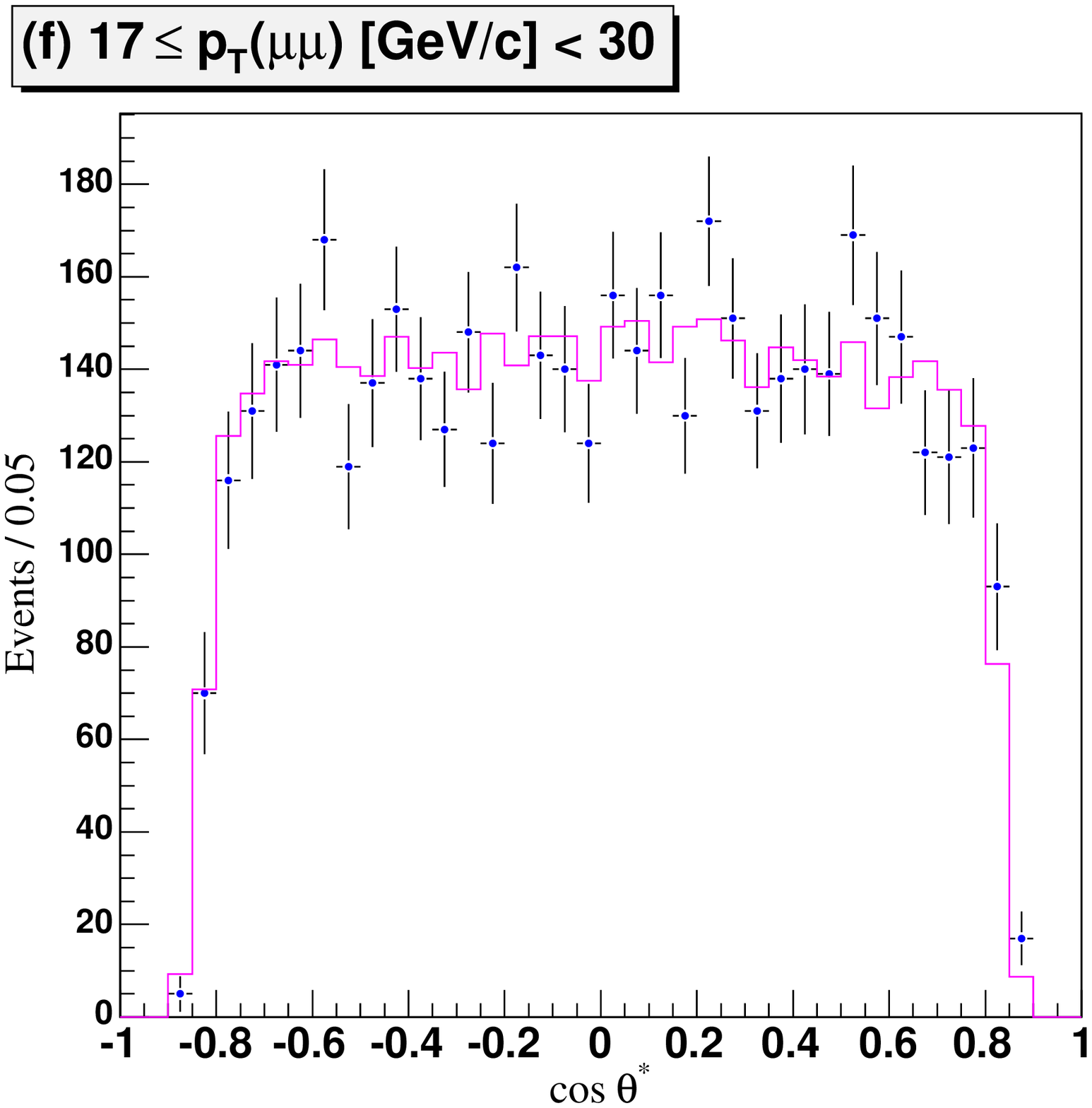}
\end{tabular}
\caption{ The measured $\cos{\theta^*}$ distribution in different
  momentum bins for prompt \prt{J/\psi} in \un{0.8}{fb^{-1}} at
  CDF. The $p_T$ bins are from left to write, top to bottom:
  \un{[5-6], [6-7], [7-9], [9-12], [12-17], [17-30]}{GeV}, A
  flat distribution indicates no polarisation, a concave ($\cup$)
  distribution transverse polarisation and a convex ($\cap$)
  distribution longitudinal polarisation.\label{fig:JpsiPolPerBin}}
\end{figure}
\begin{figure*}
\mbox{\includegraphics[height=0.3\textwidth]{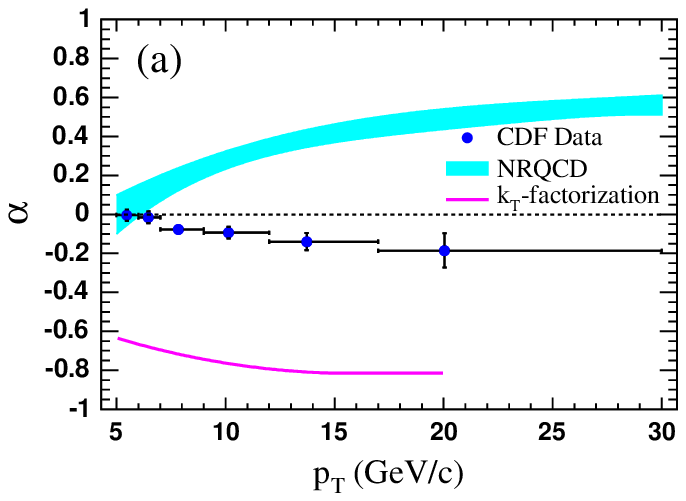}
 \hspace{2em}
 \includegraphics[height=0.3\textwidth]{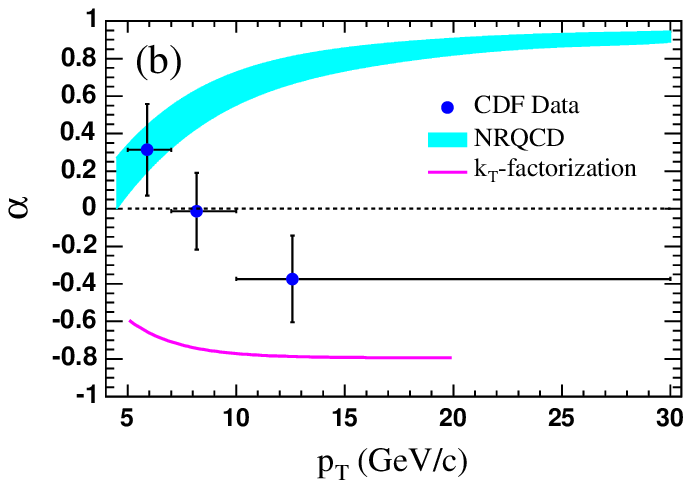}}
\caption{The polarisation parameter $\alpha$ vs momentum for prompt
  \prt{J/\psi}(a) and prompt \prt{\psi(2S)} (b) at CDF
  \jcite{Abulencia:2006_JpsiPol}.  Negative $\alpha$ correspond to
  longitudinal polarisation, positive $\alpha$ to transverse
  polarisation. The NRQCD calculation \cite{Cho:1994ih, Beneke:1995yb,
  Braaten:1999qk} is superimposed in turquoise, the $k_{T}$
  factorisation calculation \cite{Baranov:2002cf} in magenta.
  \label{fig:alphaJpsiPrompt}}
\end{figure*}
\begin{figure}
 \includegraphics[width=0.9\columnwidth]{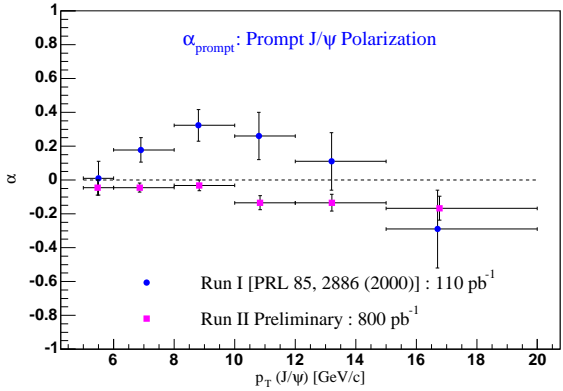}
\caption{The polarisation parameter $\alpha$ vs momentum for prompt
  \prt{J/\psi} at CDF Run~I and Run~II, where the Run~II data were
  re-analysed to match the binning in Run~I. \label{fig:alphaJpsiPrompt_run1_run2}}
\end{figure}
  In \un{0.8}{fb^{-1}}, CDF finds $0.8M$ prompt \prt{J/\psi \to \mu^+
  \mu^-} decays. The mass distribution of the \un{\mu^+\mu^-} pairs is
  shown in Fig \ref{fig:JpsiMass}.  \prt{J/\psi} originating from
  \prt{B} decays are rejected using an impact parameter cut. The
  $\cos\theta^*$ distributions are analysed in six bins of different
  $p_T$, as shown in Fig~\ref{fig:JpsiPolPerBin}. The result in terms
  of the parameter $\alpha$ as a function of $p_T$ is shown in
  Fig~\ref{fig:alphaJpsiPrompt} (a). The plot shows $p_T$-dependent,
  longitudinal polarisation of the prompt \prt{J/\psi}. This
  contradicts NRQCD which predicts increasingly transverse
  polarisation with higher momenta; $k_T$ factorisation on the other
  hand appears to over-estimate the longitudinal polarisation. The
  result also disagrees with Run~I results for \un{110}{pb^{-1}} at a
  centre-of-mass energy of \un{1.8}{TeV}, where evidence of positive
  polarisation of the \prt{J/\psi} was seen. To facilitate a
  comparison between the Run~I and Run~II results, the Run~II results
  have been re-analysed using the same binning as in Run~I. The
  comparison is shown in Fig~\ref{fig:alphaJpsiPrompt_run1_run2}.

  The polarisation has also been measured in the theoretically cleaner
  (no feed down from higher states) \prt{\psi^{\prime}} channel; the
  result is shown in Fig~\ref{fig:alphaJpsiPrompt} (b). The event
  numbers are much lower than for the \prt{J/\psi} but the results
  indicate a trend for the longitudinal polarisation fraction to
  increase with higher $p_T$, inconsistent with the NRQCD calculation.
  These results have been published in \cite{Abulencia:2006_JpsiPol}.

\section{Measurement of the \prt{\bf \Upsilon} Polarisation}
  The theoretical methods describing the production and polarisation
  of \prt{\Upsilon(1S)} and \prt{\Upsilon(2S)} in $p-\bar{p}$
  collisions are equivalent to those for charmonium presented in the
  previous section.

\begin{figure}
  \includegraphics[width=0.9\columnwidth]{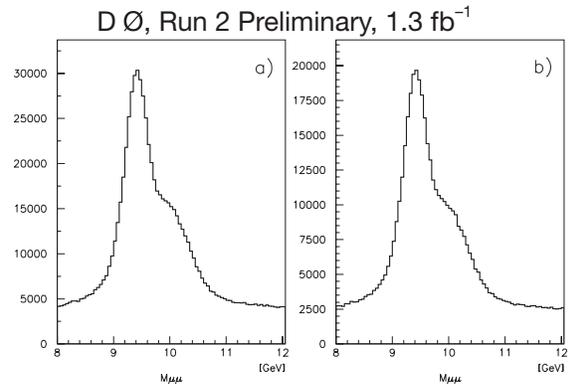}
\caption{The $\mu^+ \mu^-$ mass distribution at \DZero\ after
  selection cuts, on the left for all events, on the right only for
  those events selected by the di-muon trigger.\label{fig:UpsilonMassAll}}
\end{figure}
\begin{figure}
  \rotatebox{0}{
\includegraphics[width=0.9\columnwidth]{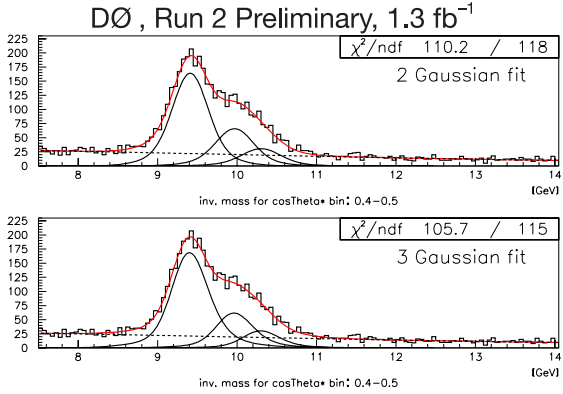}}
\caption{The $\mu^+ \mu^-$ mass distribution at \DZero\ after
  selection cuts, for one bin in $\cos|\theta^*|$ and $p_T$, with
  $|\cos\theta^*| \in [0.4,0.5]$ and $p_T \in [\un{10}{GeV},
  \un{15}{GeV}]$.  The plot shows the total fit to the data in red for
  the two fit models briefly described in the text, and the Gaussians
  describing the individual contribution of the \prt{\Upsilon(1S)},
  \prt{\Upsilon(2S)} and \prt{\Upsilon(3S)}. The background fit is
  shown as a dotted line.\label{fig:UpsilonMassBin}}
\end{figure}
  The \DZero\ experiment measures the polarisation of the
  \prt{\Upsilon(1S)} and the \prt{\Upsilon(2S)} as a function of the
  \prt{\Upsilon} transverse momentum in \un{1.3}{fb^{-1}} of data in
  Tevatron Run~II. Data are selected in the di-muon channel. To
  achieve a more reliable estimate of the trigger efficiency in the
  cross section calculation, only events selected by the di-muon
  trigger are used in the analysis. The invariant mass of the $\mu^+
  \mu^-$ pairs near the \prt{\Upsilon} mass is shown in
  Fig~\ref{fig:UpsilonMassAll} for all data passing the selection
  cut (a), and those that also were selected by the di-muon
  trigger (b). Figure~\ref{fig:UpsilonMassBin} shows the di-muon mass
  spectrum for one bin in $|\cos\theta^*|$ and $p_T$, with the fit to
  the data superimposed, and the Gaussians describing the individual
  contribution of the \prt{\Upsilon(1S)}, \prt{\Upsilon(2S)} and
  \prt{\Upsilon(3S)}. Because of the relatively small number of
  \prt{\Upsilon(3S)} events, to ensure a stable fit, the width,
  relative position and the fraction of events in the
  \prt{\Upsilon(3S)} peak were taken from MC simulation and fixed in
  the fit. In another fit, the position of the \prt{\Upsilon(3S)} peak
  is allowed to float. The difference between the two approaches is
  taken as a systematic error.
\begin{figure}
\includegraphics[width=0.9\columnwidth]{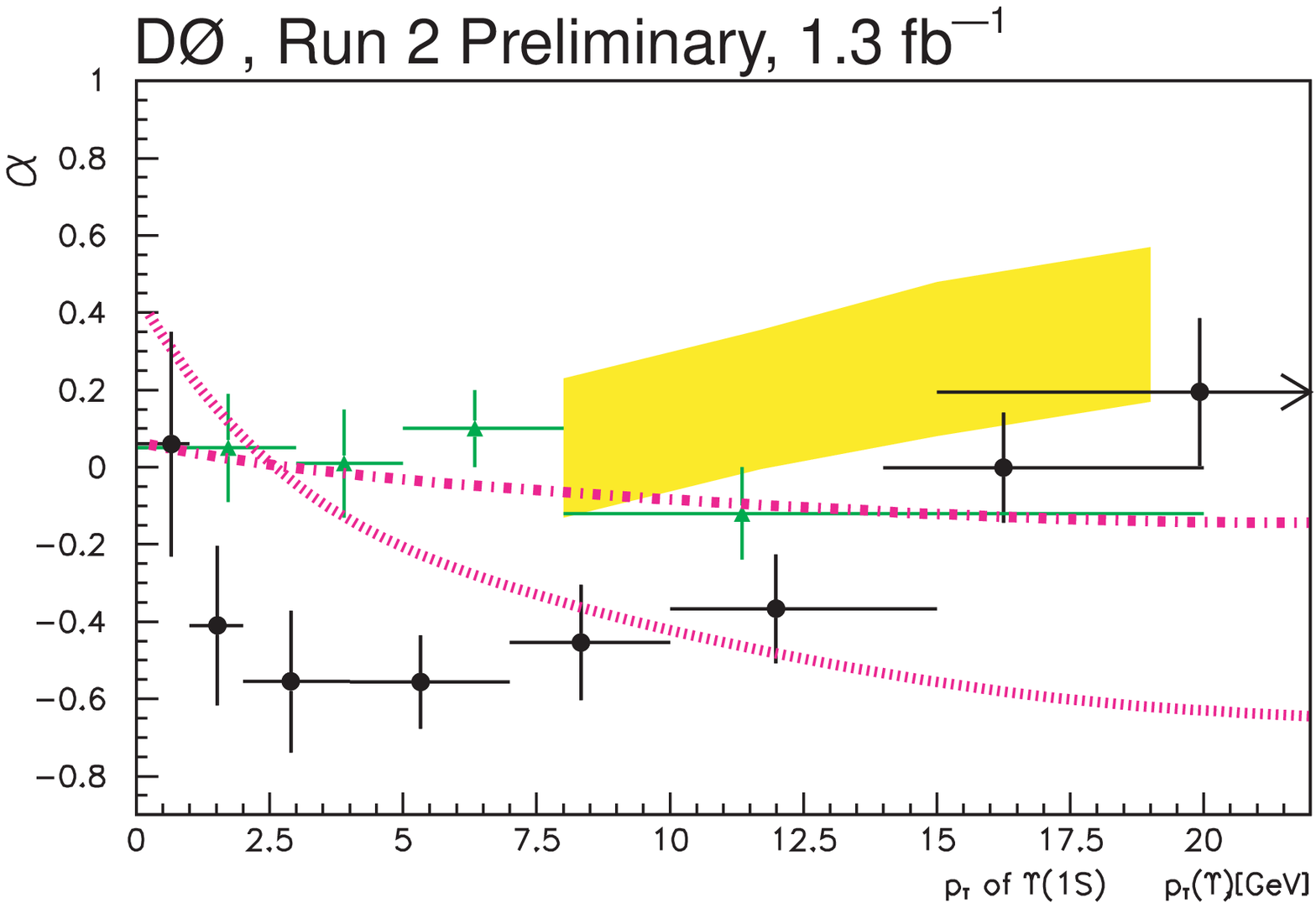}
\caption{The polarisation parameter $\alpha$ versus $p_T$ measured in
  \prt{\Upsilon(1S)\to \mu \mu} events in \un{1.3}{fb^{-1}} of data at
  \DZero. The measured results in eight $p_T$ bins are shown as black
  circles with error bars. The yellow band corresponds the NRQCD
  prediction given in \cite{Braaten:2000gw}. The magenta lines
  correspond to two limit cases of the $k_T$-factorisation model
  \jcite{Baranov:2002cf, Baranov:2007ay}. The CDF Run~I result \jcite{Acosta:2001gv} is
  shown as the green triangles with error
  bars.\label{fig:Upsilon1SPol}}
\end{figure}
\begin{figure}
\includegraphics[width=0.9\columnwidth]{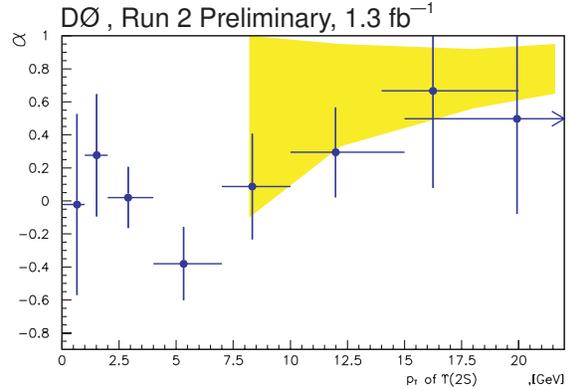}
\caption{The polarisation parameter $\alpha$ versus $p_T$ measured in
  $\Upsilon(2S)\to \mu \mu$ events in \un{1.3}{fb^{-1}} of data at
  \DZero. The data are shown as black circles with error bars. The
  yellow band shows the NRQCD prediction from
  \cite{Braaten:2000gw}.\label{fig:Upsilon2SPol}}
\end{figure}
 The \prt{\Upsilon(1S)} polarisation vs $p_T$ is shown in
 Fig~\ref{fig:Upsilon1SPol}. The data represent the result for an
 admixture of directly produced \prt{\Upsilon(1S)} and
 \prt{\Upsilon(1S)} from other sources, in particular from
 \prt{\Upsilon(2S)}, \prt{\Upsilon(3S)} and \prt{\chi_b}.  For
 comparison, the CDF Run~I result is superimposed in the same plot, as
 well as theory predictions for the polarisation parameter $\alpha$ of
 this admixture. The NRQCD prediction given in \cite{Braaten:2000gw}
 is shown as a yellow band, and two limit-cases of the
 $k_T$-factorisation model \cite{Baranov:2002cf, Baranov:2007ay} are
 given as magenta curves; the flatter of the two magenta curves
 corresponds to $k_T$-factorisation with the assumption of full
 quark-spin conservation, the steeper curve to full quark-spin
 depolarisation.  \DZero\ find clear evidence for longitudinal
 polarisation, in contradiction to the NRQCD calculation. The
 polarisation vs $p_T$ in the statistically less powerful
 \prt{\Upsilon(2S)} channel, which includes contributions from
 directly produced \prt{\Upsilon(2S)} as well as feed-down from
 \prt{\Upsilon(3S)} and \prt{\chi_b(2P)}, is shown in
 Fig~\ref{fig:Upsilon2SPol}. In contrast to the result for the
 \prt{\Upsilon(1S)}, the measured \prt{\Upsilon(2S)} polarisation is
 compatible with NRQCD.

\section{Relative Production Cross Section \prt{\bf \chi_{c2}} and
 \prt{\bf \chi_{c1}}}

 The measurement of the \prt{\chi_{cJ}} ($\mathrm{J = 1, 2}$) cross
 section is an interesting measurement in its own right, as well as
 important input to \prt{J/\psi} production measurements to which it
 provides an important source of feed-down. Experimentally, this
 measurement has always been plagued by the poor mass resolution of
 the reconstructed \prt{\chi_{cJ}} which is due to the soft photon in
 the decay chain \prt{\chi_{cJ} \to J/\psi \gamma}. The high
 luminosity at the Tevatron now provides sufficient statistics to
 restrict the analysis only to events where the photon undergoes
 conversions to an \prt{e^+ e^-} pair. The momenta of the electrons
 can be measured precisely resulting in a far superior energy
 resolution of the photon than would be possible by measuring the
 photon's energy in the calorimeter. This leads to an excellent
 resolution of the reconstructed \prt{\chi_{cJ}} mass, allowing a
 separation of the $\mathrm{J=0,1,2}$ states.
 \begin{figure}
  \includegraphics[width=0.9\columnwidth]{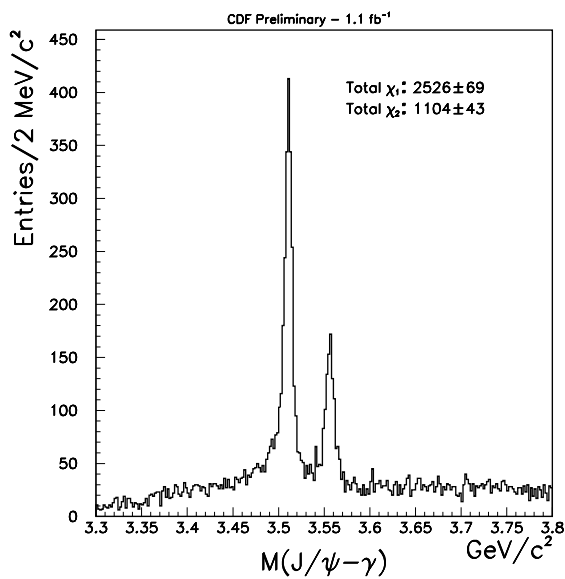}
\caption{Reconstructed \prt{\mu^+ \mu^- (\gamma\to e^+ e^-)} mass,
  showing clearly two distinct mass peaks at the \prt{\chi_{c1}} and
  \prt{\chi_{c2}} mass.\label{fig:chiMass}}
 \end{figure}
 CDF reconstructs \prt{\chi_{cJ}} in in the channel \prt{\chi_{cJ} \to
 J/\psi(\mu \mu) \gamma(e^+ e^-)} within the kinematic range
 \un{p_T(J/\psi) \in \un{[4, 20]}{GeV}}. To select photon conversions,
 the \prt{e^+ e^-} pair is a required to form a well-reconstructed
 vertex a large distance (\un{>12}{cm}) from the beam, well inside the
 instrumented region of CDF.  The spectrum of the \prt{(\mu^+ \mu^-
 \gamma)} invariant mass is shown in Fig~\ref{fig:chiMass}, showing
 two well-separated peaks at the \prt{\chi_{c1}} mass and the
 \prt{\chi_{c2}} mass. No significant evidence for \prt{\chi_{c0}}
 production can be seen. After all selection cut total number of
 \prt{\chi_{cJ} \to J/\psi (\mu \mu)\; \gamma(e^+ e^-)} events found in
 \un{1.1}{fb^{-1}} is $\sim 7k$.

 The distance between the beamline and the $\mu^+ \mu^-$ vertex is
 used to separate the prompt contribution from \prt{\chi_{cJ}}
 originating from \prt{B} decays. The measured ratio for prompt
 production is:
\[
 \frac{\sigma(\chi_{c2})}{\sigma(\chi_{c1})} 
= 0.70 \pm 0.04\mathrm{(stat)} \pm 0.03\mathrm{(sys)} \pm 0.06\mathrm{(BF)}
\]
 for $p_T(\prt{J/\psi}) \in \un{[4, 20]}{GeV}$. The last contribution
 to the error is due to the uncertainty in the branching fraction of
 \prt{\chi_{cJ} \to J/\psi \gamma}.  This result is at odds with the
 expectation from the colour octet model, which, by counting
 spin-states, predicts a ratio of $5/3$.

\section{Conclusions}

 The large charm and bottom production cross section at
 \un{\sqrt{s}=1.96}{TeV} proton-antiproton collisions, combined with
 the capabilities of the \DZero\ and the CDF detector provide the
 opportunity for new measurements with unprecedented statistics and
 precision.

 In this paper we presented Run~II results on single and correlated
 open charm and bottom production as well as quarkonium production and
 polarisation. The same theoretical framework that managed to describe
 successfully the unexpectedly large charmonium production cross
 observed at Tevatron Run~I now fails to account for the significant
 longitudinal polarisation of charmonium and bottomium observed in
 Run~II, nor the \prt{\chi_{c2}} and \prt{\chi_{c1}} cross section
 ratio. Several alternative models are being developed, but at the
 time of writing this paper, none has provided a detailed quantitative
 post-diction of the charmonium polarisation vs $p_T$ that matches the
 observed data.

 So heavy flavour production is, refreshingly, a field that is clearly
 led by experiment. In response to the data from the Tevatron, some
 of which have been presented here, we can look forward to new
 calculations and models, offering new descriptions of the mechanism
 of heavy flavour, and in particular quarkonium, production. We will
 also see further new measurements making use of the large amount of
 data yet to be analysed. Only about $1/3$ of the Tevatron data taken
 so far have been used for the results presented here, and with the
 machine going stronger than ever, we can expect to at least double
 the integrated Tevatron luminosity before the end of Run~II.

\bigskip 

\bibliography{bibliography}


\end{document}